\documentclass[article, 10pt]{article}

\usepackage{bm}
\usepackage{geometry}
\usepackage{abstract}
\usepackage{lmodern}
\usepackage{amsmath}
\usepackage{graphicx}
\usepackage{booktabs}
\usepackage{authblk}
\usepackage[colorlinks=true, allcolors=blue]{hyperref}
\usepackage[english]{babel}
\usepackage{rotating}
\usepackage{amssymb}
\usepackage{doi}
\usepackage{hyperref}

\usepackage[backend=biber,
            style=numeric-comp,
            sorting=none,
            doi=true,
            url=false,
            giveninits=true,
            maxnames=10,
            minnames=1]{biblatex}
\renewbibmacro{in:}{} 
\AtEveryBibitem{
  \clearfield{issn}
  \clearfield{isbn}
  \clearfield{eprint}
  \clearfield{number}
}

\DeclareNameAlias{default}{last-first}
\DeclareNameFormat{given-name}{#1}
\DeclareNameFormat{given-family}{#1}

\DeclareFieldFormat[article]{volume}{\textbf{#1}}

\usepackage{xr} 
\begin{document}

\title{Multi-fidelity learning for interatomic potentials: Low-level forces and high-level energies are all you need} 

\author[1,2]{Mitchell Messerly}
\author[1]{Sakib Matin}
\author[1,3,4]{Alice E. A. Allen}
\author[1]{Benjamin Nebgen}
\author[1,3]{Kipton Barros}
\author[5]{Justin S. Smith}
\author[6]{Nicholas Lubbers}
\author[1,7]{Richard Messerly\thanks{Corresponding author: messerlyra@ornl.gov}}

\affil[1]{Theoretical Division, Los Alamos National Laboratory, Los Alamos, NM 87545, USA}
\affil[2]{Department of Mechanical Engineering, Brigham Young University, Provo, UT 84604, USA}
\affil[3]{Center for Nonlinear Studies, Los Alamos National Laboratory, Los Alamos, New Mexico 87545, United States}
\affil[4]{Max Planck Institute for Polymer Research, Ackermannweg 10, 55128 Mainz, Germany}
\affil[5]{Nvidia Corporation, Santa Clara, CA 95051, United States}
\affil[6]{Computer, Computational, and Statistical Sciences Division, Los Alamos National Laboratory, Los Alamos, New Mexico 87545, United States}
\affil[7]{National Center for Computational Sciences Division, Oak Ridge National Laboratory, Oak Ridge, TN 37830, USA}

\date{This manuscript has been authored in part by UT-Battelle, LLC, under contract DE-AC05-00OR22725 with the U.S. Department of Energy (DOE).  The U.S. government retains and the publisher, by accepting the article for publication, acknowledges that the U.S. government retains a nonexclusive, paid-up, irrevocable, worldwide license to publish or reproduce the published form of this manuscript, or allow others to do so, for U.S. government purposes.  DOE will provide public access to these results of federally sponsored research in accordance with the DOE Public Access Plan (http://energy.gov/downloads/doe-public-access-plan).} 
\maketitle

\clearpage
\newpage

\section*{Abstract}
The promise of machine learning interatomic potentials (MLIPs) has led to an abundance of public quantum mechanical (QM) training datasets. The quality of an MLIP is directly limited by the accuracy of the energies and atomic forces in the training dataset. Unfortunately, most of these datasets are computed with relatively low-accuracy QM methods, e.g., density functional theory with a moderate basis set. Due to the increased computational cost of more accurate QM methods, e.g., coupled-cluster theory with a complete basis set extrapolation, most high-accuracy datasets are much smaller and often do not contain atomic forces. The lack of high-accuracy atomic forces is quite troubling, as training with force data greatly improves the stability and quality of the MLIP compared to training to energy alone. Because most datasets are computed with a unique level of theory, traditional single-fidelity learning is not capable of leveraging the vast amounts of published QM data. In this study, we apply multi-fidelity learning to train an MLIP to multiple QM datasets of different levels of accuracy, i.e., levels of fidelity. Specifically, we perform three test cases to demonstrate that multi-fidelity learning with both low-level forces and high-level energies yields an extremely accurate MLIP---far more accurate than a single-fidelity MLIP trained solely to high-level energies and almost as accurate as a single-fidelity MLIP trained directly to high-level energies and forces. Therefore, multi-fidelity learning greatly alleviates the need for generating large and expensive datasets containing high-accuracy atomic forces and allows for more effective training to existing high-accuracy energy-only datasets. Indeed, low-accuracy atomic forces and high-accuracy energies are all that are needed to achieve a high-accuracy MLIP with multi-fidelity learning.

\section{Introduction}

Atomistic simulations, e.g., molecular dynamics (MD), find applications in many fields of science, including biology, materials science, and chemistry. The predictive capability of MD depends primarily on the accuracy of the forces acting on each atom. Historically, these forces have been calculated using either quantum mechanical (QM) methods, such as density functional theory (DFT), or classical mechanical methods, such as force fields (FFs), also referred to as interatomic potentials. Both QM and FF methods have limitations. Although QM calculations are very accurate, they are also extremely expensive, with a cost that increases rapidly with the number of electrons (typically to at least the third power). In comparison with QM, FFs are significantly less expensive and typically scale linearly with the number of atoms, rather than with the number of electrons. However, FFs are less accurate than QM because they use inflexible functional forms with relatively few parameters that are fit to limited amounts of data. Recently, machine learning interatomic potentials (MLIPs),~\cite{thiemann_introduction_2024} and especially neural network-based MLIPs,~\cite{martin-barrios_overview_2024} promise a ``best-of-both-worlds'' scenario --- computational costs and scaling similar to classical FFs but accuracy similar to the QM method used to generate the training data.~\cite{kulichenko2021rise}  

While several different MLIP architectures exist,~\cite{behler2011atom,smith2017ani,schutt_schnet_2017,e3equivariant,anstine2024aimnet2,maceoff23,hedelius_triptransformer_2024} the overall quality of an MLIP, both in terms of accuracy and robustness,~\cite{unke2021machine} is primarily predicated on the training dataset.~\cite{miksch_strategies_2021} By contrast to classical FFs, which are typically parameterized with limited amounts of experimental data, training datasets for MLIPs consist of atomic positions with their corresponding energies and/or forces computed with the desired QM method. Because neural networks are extremely flexible ``universal function approximators'', developing reliable and transferable neural network MLIPs requires enormous amounts of training data.~\cite{gokcan2022} For this reason, many studies focus on developing large and diverse QM datasets.~\cite{Smith2017b,Ramakrishnan2014QuantumMolecules,hoja2021qm7,smith2020ani,schreiner2022transition1x,mathiasen_generating_2023,barroso-luque_open_2024,levine_open_2025}

While the rapid growth in freely available training datasets is extremely beneficial to the greater scientific community,~\cite{kulichenko2024data} several challenges currently limit the practical usefulness of the plethora of public QM data. One challenge is incompatibility across different datasets. In order to have a smooth MLIP, all energies and/or forces need to be computed with precisely the same QM method (e.g., the same functional, the same basis set, the same energy cut-off). Even calculations performed with the same QM method but with two different codes (or even with two different versions of the same code) can result in non-negligible differences in the energies and/or forces.~\cite{schreiner_neuralnebneural_2022} Because different datasets are not all computed with the same QM level of theory or the same QM software, naively combining public datasets for training an MLIP is extremely risky. Thus, although there are many published datasets, researchers commonly generate a completely new training dataset for the specific system of interest using their preferred QM method and code. 

Another challenge limiting the impact of public training data is that most datasets are computed using relatively low-level QM methods, with DFT being the most common method of choice.~\cite{kulichenko2024data} Although modern functionals can achieve reasonable accuracy given a large enough basis set, most QM datasets are computed with traditional functionals and moderate basis sets. Unfortunately, highly accurate QM methods (e.g., coupled-cluster, CC,~\cite{bartlett2024} with a complete basis set, CBS, extrapolation)~\cite{petersson1988} are orders of magnitude more computationally expensive. Consequently, the few datasets computed with high-level QM methods are much smaller in terms of both number of calculations and size of systems (i.e., number of atoms).~\cite{smith2020ani} 

A related challenge is that several QM codes cannot provide analytical forces for the most accurate QM methods.~\cite{orca,g16,nwchem} Although most QM codes can compute numerical forces at any level of theory, these forces are more approximate and more expensive to compute as they require performing numerous energy calculations at slightly perturbed atomic configurations. For these reasons, although some recent efforts have expended significant computational resources to calculate large amounts of CC-level forces,~\cite{allen_reactive_2025} most datasets computed with high-level QM methods do not contain force data.~\cite{smith2020ani,donchev_quantum_2021,daru_coupled_2022,hu2023neural,qu_breaking_2021}  For example, while the ANI-1x dataset contains DFT-level energies and forces for $\approx$4.5M configurations of small organic molecules \textit{in vacuo}, the ANI-1ccx dataset contains only CC-level energies for a subset of $\approx$460k configurations.~\cite{smith2020ani} 

Force data are especially valuable for MLIP development. Because forces are vector quantities associated with each atom, a single QM calculation returns one system energy but 3$N$ force components, where $N$ is the number of atoms. Forces are also a local property, whereas system energy is a global property. Thus, force data facilitate more direct learning of the relationship between the local atomic environment and the potential energy surface. Furthermore, as forces are the negative gradient of the energy, training to forces helps ensure a ``smooth'' potential. For these reasons, inclusion of force data~\cite{Smith2020SimpleAE} (and, by extension, Hessian data)~\cite{rodriguez_does_2025,amin_towards_2025} is extremely advantageous for training speed, MLIP accuracy, and MLIP stability in MD simulations. For example, due to both the lack of force data in ANI-1ccx and the fewer number of molecular structures in ANI-1ccx compared to ANI-1x, previous MLIPs trained solely to the ANI-1ccx dataset were more limited in accuracy despite being trained to more expensive CC-level energy data.~\cite{smith2019approaching,chigaev_lightweight_2023,zaverkin2023transfer}  

\begin{figure}[htb!]
    \includegraphics[width=4.2in]{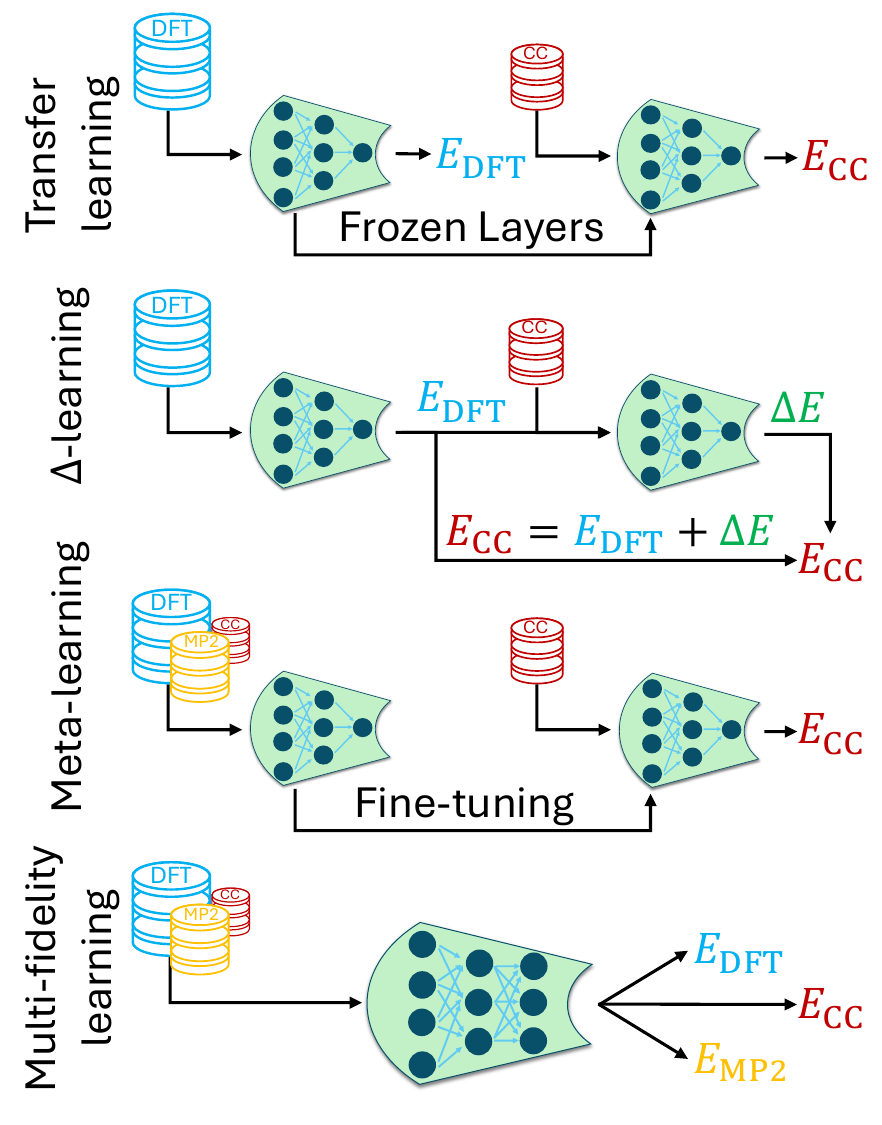}
    \caption{Comparison of different methods capable of training to multiple datasets at different levels of theory. Although meta-learning and multi-fidelity learning can utilize many datasets, for visual simplicity, only three levels of theory are depicted. Low-level data are represented with density functional theory (DFT). Intermediate-level data are represented with M{\o}ller-Plesset second-order perturbation theory (MP2). High-level data are represented with coupled-cluster theory (CC). While some variation exists in the literature as to the implementation of each method, the depictions in this figure are consistent with the transfer learning and $\Delta$-learning work of Smith et al.,~\cite{smith2019approaching} the meta-learning work of Allen et al.,~\cite{allen_learning_2024} and the multi-fidelity learning work of Jacobson et al.~\cite{jacobson_leveraging_2023}}
    \label{fig:Multidataset_Overview}
\end{figure}

The abundance of incompatible low-accuracy datasets and the scarcity of high-accuracy forces requires more sophisticated training techniques that can simultaneously learn from multiple datasets. Figure~\ref{fig:Multidataset_Overview} compares four such multi-dataset training paradigms, namely, transfer learning (TL),~\cite{hu2023neural,smith2019approaching,zaverkin2023transfer,stippell_building_2024} $\Delta$-learning ($\Delta$L),~\cite{daru_coupled_2022,qu_breaking_2021,ramakrishnan_big_2015,Nandi2024JCTC,burrill_mltb_2025} meta-learning ($\mu$L),~\cite{allen_learning_2024} and multi-fidelity learning (MFL).~\cite{jacobson_leveraging_2023,cui_multi-fidelity_2025} All four methods present unique advantages and disadvantages when training to multiple datasets with different levels of theory. Both TL and $\Delta$L are powerful methods in the limit of only two datasets at different levels of theory, whereas $\mu$L and MFL are theoretically amenable to any number of levels of theory. Both TL and $\Delta$L consist of first training an MLIP on the large, low-accuracy dataset, followed by training to the smaller, high-accuracy dataset. By contrast, $\mu$L and MFL train to all datasets simultaneously, although $\mu$L also requires a subsequent fine-tuning step. Thus, TL, $\Delta$L, and $\mu$L are two-step methods that result in separate MLIPs for each level of theory. By contrast, MFL is a single-step method which trains a single MLIP that predicts multiple energies, with one output energy for each level of accuracy (i.e., level of fidelity). MFL is also often referred to as multi-level learning,~\cite{nandi_multixc-qm9_2023,Heinen_2024} multi-head learning, and multi-task learning~\cite{tang_approaching_2024,Unke2019PhysNet:Charges} (see Appendix for discussion of terminology).

Previous studies demonstrate that TL,~\cite{smith2019approaching} $\Delta$L,~\cite{smith2019approaching} and $\mu$L~\cite{allen_learning_2024} achieve significant improvement when training to two datasets that overlap perfectly in configuration space, namely, the ANI-1x and ANI-1ccx datasets. Specifically, all three multi-dataset methods report approximately 20\% higher accuracy compared to training solely to the ANI-1ccx CC-level energies. By contrast, previous work did not see significant improvement when leveraging MFL to simultaneously train to ANI-1x and ANI-1ccx.~\cite{jacobson_leveraging_2023} However, this previous study did not include the DFT-level atomic forces when performing MFL, only the DFT-level energies and CC-level energies were used for training. The justification for training only to energies was to avoid biasing the MLIP towards the ANI-1x DFT-level forces, since ANI-1ccx does not contain CC-level force data. Consequently, MFL with just DFT-level energies and CC-level energies achieved comparable accuracy to training solely to CC-level energies. Additional studies also tested the limits of MFL by training to several different levels of theory simultaneously.~\cite{jacobson_leveraging_2023,nandi_multixc-qm9_2023,shoghi_molecules_2023,pasini_scalable_2024} Learning from multiple datasets covering different regions of chemical space offers clear benefits for improved transferability, but does not necessarily lead to improved accuracy. 

The primary goal of this work is to demonstrate that MFL can result in an MLIP with extremely accurate high-level energies and forces, even when high-level forces are not available for training. Our hypothesis is that high-level forces can be learned implicitly with MFL by simultaneously training to low-level forces and high-level energies. To validate our hypothesis, we apply MFL to train an MLIP to the DFT-level energies and forces and CC-level energies for the $\approx$460k structures common to both the ANI-1x and ANI-1ccx datasets. These two levels of theory were chosen to enable comparison with previous studies that trained to the same datasets but utilized TL,~\cite{smith2019approaching} $\Delta$L,~\cite{smith2019approaching} $\mu$L,~\cite{allen_learning_2024} and MFL without DFT-level forces.~\cite{jacobson_leveraging_2023} For further validation, we repeat this MFL process but with two other intermediate-level QM methods in place of the CC-level energy data. For completeness, we briefly investigate any potential benefit of training with force data from more than two levels of theory. 

\section{Methods}

All models in this study utilize the ``Hierarchically Interacting Particle Neural Network'' (HIP-NN) architecture, a state-of-the-art message-passing graph-based neural network MLIP.~\cite{lubbers2018hierarchical} A detailed overview of HIP-NN is provided in the Appendix Section. The HIP-NN architecture is fundamentally different from the ``Accurate NeurAl networK engINe for Molecular Energies'' (ANAKIN-ME, or ANI)~\cite{smith2017ani} feed forward neural network architecture used in previous MFL work.~\cite{jacobson_leveraging_2023} Although our primary conclusions regarding the performance of MFL do not depend on the MLIP architecture chosen, this section describes some MFL design decisions specific to HIP-NN.


\subsection{Multi-fidelity learning with HIP-NN}

\begin{figure}[htb!]
    \includegraphics[width=4in]{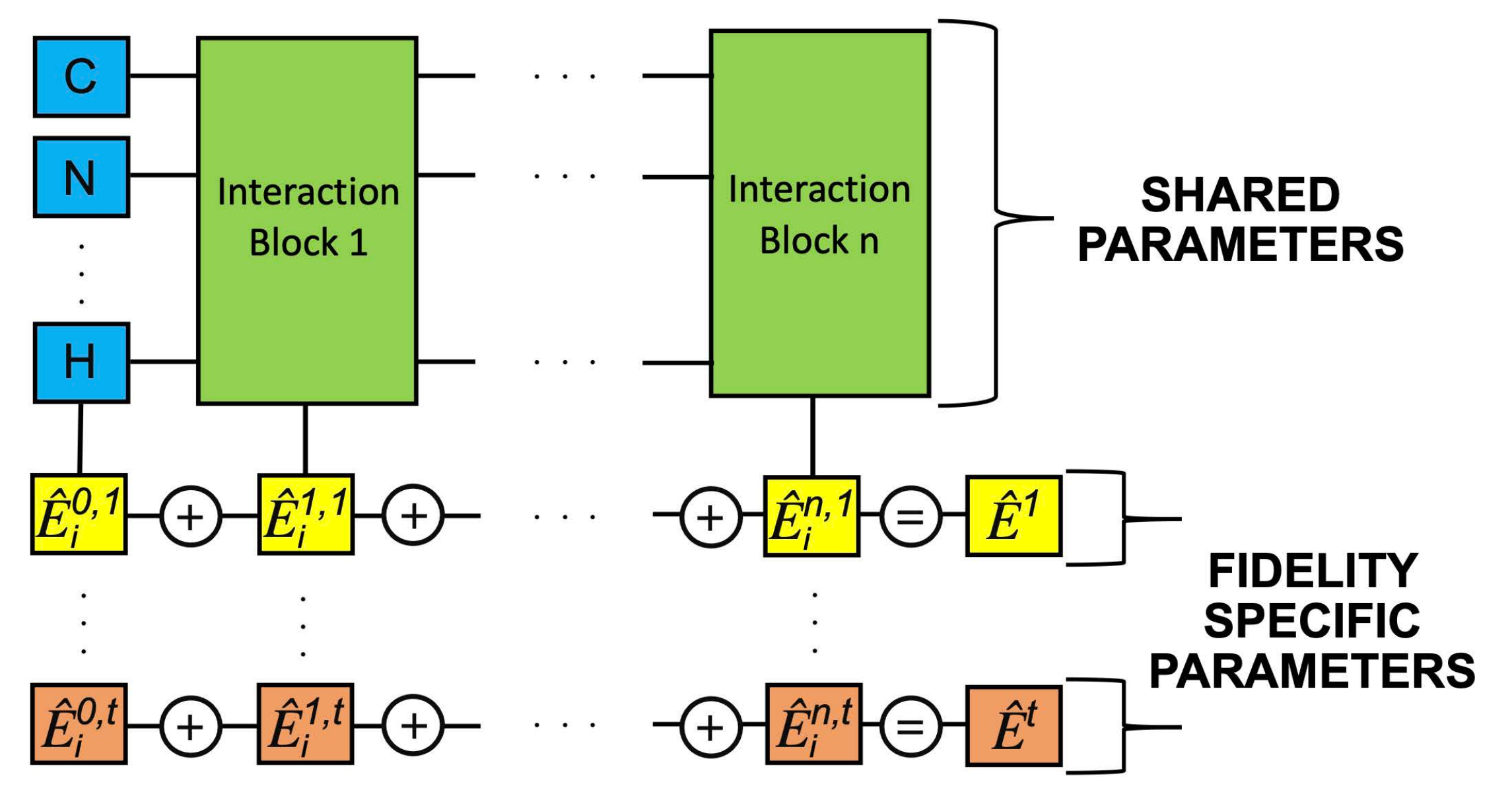}
    \caption{HIP-NN multi-fidelity architecture with shared parameters for interaction blocks (green) and fidelity-specific non-shared parameters for linear output layers (yellow and orange). $n$ refers to the interaction block and hierarchical order. $t$ refers to the level of theory. $i$ refers to the atom number.}
    \label{fig:Multi-fidelity HIPNN architecture}
\end{figure}

Although HIP-NN has never been used previously to predict energy at multiple levels of theory, standard HIP-NN is already capable of predicting multiple outputs. Therefore, no modifications to the HIP-NN code were required to perform MFL. However, training HIP-NN to multiple levels of theory requires several important design decisions. The most important decision is which layers use shared parameters (i.e., parameters trained simultaneously to multiple levels of theory) and which layers use non-shared parameters (i.e., fidelity-specific parameters trained to a single level of theory). In this work, all interaction blocks use shared parameters while all linear output layers use fidelity-specific non-shared parameters (see Figure~\ref{fig:Multi-fidelity HIPNN architecture}). Therefore, the weights and biases in Equation~\ref{int_layer} and Equation~\ref{atomic_layer} are shared parameters while the weights and biases in Equation~\ref{linear_output_layer} are the only non-shared parameters (see Appendix). 

For multi-fidelity learning, the total loss function $(L)$ includes a separate loss term for energy and forces for each level of theory, according to:
\begin{equation}
    L = \sum_t(w_{E,t} L_{E,t} + w_{F,t} L_{F,t} + w_{H,t} L_{H,t})
\end{equation}
where $t$ denotes the level of theory, $E$ stands for energy, $F$ stands for forces, $L_{E,t}$, $L_{F,t}$, and $L_{H,t}$ are the energy, force, and hierarchicality loss terms for the $t^{\rm th}$ level of theory, respectively, and $w_{E,t}$, $w_{F,t}$, $w_{H,t}$ are the weights for the energy, force, and hierarchicality loss terms for the $t^{\rm th}$ level of theory, respectively. The loss function for single-fidelity learning is obtained when the summation is over one level of theory. In this work, the weights for all contributions are equal, i.e., $w_{E,t} = 1$, $w_{F,t} = 1$, and $w_{H,t} = 1$. However, force weights are zero $(w_{F,t} = 0)$ if not training to forces from that level of theory. Several other studies also use an equal energy-to-force weight ratio when training diverse MLIP architectures to the ANI-1x dataset.~\cite{hedelius_learning_2022,zugec2024global,simeon2024tensornet,zhang2025} Although the present work utilizes constant weights throughout training, future work could investigate whether an adaptive loss weighting approach is beneficial for MFL.~\cite{ocampo_adaptive_2024} Specifically, the low-level forces would be assigned greater weights at early epochs while the high-level energies would be assigned greater weights at later epochs, similar to recent work with knowledge distillation.~\cite{matin_ensemble_2025}

The $L_{H,t}$ term serves as a regularization term that ensures the MLIP obeys the principle of hierarchicality, namely, that the hierarchical energy contribution to the total energy decreases for increasing hierarchicality, i.e., $E^{n+1}<E^n$ (see Figure~\ref{fig:Multi-fidelity HIPNN architecture} and Appendix for details). 

In this work, the energy and force loss terms are defined as
\begin{equation}
    L_E = RMSE_E + MAE_E
\end{equation}
\begin{equation}
    L_F = RMSE_F + MAE_F
\end{equation}
where $RMSE$ is the root-mean-square error and $MAE$ is the mean absolute error.  

\subsection{Datasets}

\subsubsection{Training}

Training is performed using a subset of the original ANI-1x dataset, which consists primarily of near-equilibrium (i.e., non-reactive) structures for stable organic molecules.~\cite{smith2020ani} While the complete ANI-1x dataset contains $\approx$4.5M molecular structures, we utilize a subset of nearly 460k molecular structures that correspond to the same structures as the ANI-1ccx dataset. Both the complete ANI-1x dataset and the ANI-1ccx subset were generated in previous work through active learning.~\cite{smith2018less} Although the full ANI-1x dataset contains low-level energies and forces for all $\approx$4.5M structures, high-level energies are available only for this reduced subset. Thus, utilizing the reduced dataset of $\approx$460k structures helps avoid the issue of unequal data volumes between different QM methods. Previous studies have trained reliable MLIPs for diverse architectures to both the complete ANI-1x dataset as well as significantly reduced subsets of ANI-1x.~\cite{chigaev_lightweight_2023,zaverkin2023transfer,hedelius_learning_2022,zugec2024global,simeon2024tensornet,zhang2025,haghighatlari_newtonnet_2022,kovacs2023evaluation} 

The reduced ANI-1x dataset contains energies and forces computed with several QM methods. To simplify our analysis, we focus on only four levels of fidelity: DFT/DZ, DFT/TZ, MP2/TZ, and CCSD(T)*/CBS. The two lowest fidelity levels (DFT/DZ and DFT/TZ) are both based on DFT using the hybrid $\omega$b97x functional~\cite{chai_systematic_2008} with either a double zeta (DZ) or triple zeta (TZ) basis set. Specifically, the lowest fidelity level (DFT/DZ) uses the polarized split-valence DZ (6-31G*) basis set~\cite{petersson1988} while the second lowest fidelity level (DFT/TZ) uses the valence TZ with two sets of polarization functions (def2-TZVPP) basis set.~\cite{weigend_balanced_2005} For brevity, we refer to these two DFT-based fidelity levels simply as DZ and TZ, respectively. The second highest fidelity level (MP2/TZ) corresponds to M{\o}ller-Plesset second-order perturbation theory~\cite{shee_regularized_2021} with the correlation-consistent polarized valence TZ (cc-pVTZ) basis set.~\cite{dunning1989,kendall1992} The highest fidelity level (CCSD(T)*/CBS) utilizes an approximate coupled-cluster (CC) single and double excitation with perturbative triple excitation~\cite{guo_2018} with a CBS extrapolation implementation~\cite{HALKIER1999437,neese2011} (see Smith et al.~\cite{smith2020ani} for details). For brevity, we refer to the two highest fidelity levels as MP2 and CC, respectively. While energies are available for nearly all $\approx$460k structures across all four fidelity levels, forces are only available for the DFT/DZ and DFT/TZ methods. 

\subsubsection{Testing}
Two types of test datasets are used to evaluate the accuracy of our trained models: in-sample and out-of-sample. In-sample test datasets are generated by randomly selecting 10\% of the $\approx$460k configurations in the reduced ANI-1x dataset. These $\approx$50k configurations form a held-out dataset that is not included in training or validation. Each in-sample test dataset uses the same fidelity level as the training data.

Out-of-sample test datasets come from the GDB10to13 portion of the COMP6 dataset,~\cite{smith2018less} which contains approximately 48k conformations of larger molecules (10–13 heavy atoms) from the GDB-11 database~\cite{fink_virtual_2005,fink_virtual_2007} and the GDB-13 database.~\cite{blum_970_2009} The GDB10to13 test dataset primarily serves as an extensibility test. Forces in the GDB10to13 test dataset are computed using MP2/TZ, while conformer energy differences $(\Delta E_{\rm conf})$ are computed with CCSD(T)*/CBS. To allow for comparison with previous work and to focus on chemically relevant conformers, only conformers within 100 kcal/mol of the minimum-energy conformer are included in the analysis of GDB10to13 forces and $\Delta E_{\rm conf}$.

An additional out-of-sample test dataset contains torsion scan energy differences $(\Delta E_{\rm tors})$ for small drug-like molecules. This test dataset contains the same geometries as Sellers et al.~\cite{sellers_comparison_2017} that were previously optimized with MP2 and the split-valence TZ with diffuse and polarization functions (6-311+G**) basis set.~\cite{krishnan1980} However, the energies in our torsion scan test dataset were previously recomputed using the same CCSD(T)*/CBS approach as the ANI-1ccx dataset. We also test our models against the CCSD(T)/CBS energies reported by Sellers et al. (see Supplementary Material), to allow for a fair comparison with other studies.~\cite{smith2019approaching,jacobson_leveraging_2023,matin_ensemble_2025,zariquiey_quantumbind-rbfe_2025} To further test the robustness of our models, we also perform relaxed torsion scans, wherein the positions of all atoms are optimized with a given model subject to the constraint of a fixed dihedral angle (see Supplementary Material).

Performance for each model is measured by the mean absolute error (MAE) and/or root-mean-square error (RMSE) of energies and atomic forces. The RMSE and MAE are reported as an average value with a corresponding 95\% confidence interval from an ensemble of eight MLIPs. Specifically, the RMSE and MAE are first computed for each individual ensemble member. Subsequently, the mean and standard deviation are computed from the eight different RMSE and MAE values. Consequently, the average RMSEs and MAEs reported are not based on ensemble-averaged energies and forces. Averaging the energy and forces over an ensemble of MLIPs smooths the potential energy surface and, thereby, significantly reduces the errors. However, by not averaging the predictions prior to computing the errors, it is possible to quantify the ensemble uncertainty in the errors themselves and, thus, determine whether two models are statistically different. All errors reported in this work correspond to test errors, i.e., the errors computed on either the held-out dataset or the GDB10to13 testset. No training or validation errors are reported. 

\subsection{Models}

A comprehensive set of training tests is performed with each level of theory for both single-fidelity (SF) and multi-fidelity (MF) machine learning interatomic potentials (MLIPs). Table~\ref{tbl:All ensembles} gives a summary of all MLIPs trained in this work, as well as the shorthand notation adopted to distinguish between models. The SF-MLIPs allow for a baseline comparison with the MF-MLIPs as well as a benchmark comparison with the published ANI MLIPs trained to the same level of theory. An ensemble consisting of eight MLIPs is trained for each model. See Supplementary Material for model and training details.

\begin{table}[htb!]
  \caption{All MLIPs trained in this work. SF = single-fidelity. MF = multi-fidelity. DZ = DFT/DZ. TZ = DFT/TZ. MP2 = MP2/TZ. CC = CCSD(T)*/CBS. E = energies. F = forces. \checkmark signifies MLIP was trained with data from this level of theory (DFT/DZ, DFT/TZ, MP2/TZ, or CCSD(T)*/CBS) and property (E or F).}
  \label{tbl:All ensembles}
  \begin{tabular}{c c c c c c c }
  \hline
     & \multicolumn{2}{c}{DFT/DZ} & \multicolumn{2}{c}{DFT/TZ} & MP2/TZ & CCSD(T)*/CBS \\
  
    Experiment Name & E & F & E & F & E & E \\
\hline
SF-DZ/E& \checkmark & -- & -- & -- & -- & -- \\
SF-DZ/EF& \checkmark & \checkmark & -- & -- & -- & -- \\
SF-TZ/E& -- & -- & \checkmark & -- & -- & -- \\
SF-TZ/EF& -- & -- & \checkmark & \checkmark & -- & -- \\
SF-MP2/E& -- & -- & -- & -- & \checkmark & -- \\
SF-CC/E& -- & -- & -- & -- & -- & \checkmark \\
MF-DZ/E-TZ/E& \checkmark & -- & \checkmark & -- & -- & -- \\
MF-DZ/EF-TZ/E& \checkmark & \checkmark & \checkmark & -- & -- & -- \\
MF-DZ/E-TZ/EF& \checkmark & -- & \checkmark & \checkmark & -- & -- \\
MF-DZ/EF-TZ/EF& \checkmark & \checkmark & \checkmark & \checkmark & -- & -- \\
MF-DZ/EF-MP2/E& \checkmark & \checkmark & -- & -- & \checkmark & -- \\
MF-TZ/EF-MP2/E& -- & -- & \checkmark & \checkmark & \checkmark & -- \\
MF-DZ/E-CC/E& \checkmark & -- & -- & -- & -- & \checkmark \\
MF-DZ/EF-CC/E& \checkmark & \checkmark & -- & -- & -- & \checkmark \\
MF-TZ/E-CC/E& -- & -- & \checkmark & -- & -- & \checkmark \\
MF-TZ/EF-CC/E& -- & -- & \checkmark & \checkmark & -- & \checkmark \\
MF-DZ/EF-TZ/EF-CC/E& \checkmark & \checkmark & \checkmark & \checkmark & -- & \checkmark \\
\hline
  \end{tabular}
\end{table}


\section{Results}

The primary objective of this study is to demonstrate that, even in the absence of high-level forces, MFL can achieve accuracy similar to SF training directly to high-level energies and forces. We perform the following three test cases to achieve this objective.
   
The first test case of MFL utilizes DFT-level energies and forces computed with two different basis sets, DZ and TZ. The larger TZ basis set provides the high-level data in this test case. Although we do not anticipate significant differences in forces for such a marginal increase in basis set size, this first test case is insightful because the reduced ANI-1x training dataset contains both low-level (DFT/DZ) and high-level (DFT/TZ) atomic forces. Thus, we are able to train an SF-MLIP directly to DFT/TZ energies and forces (SF-TZ/EF). The SF-TZ/EF model serves as a useful benchmark for comparison with the MF-MLIP trained without DFT/TZ forces (MF-DZ/EF-TZ/E).

The second test case of MFL also utilizes DFT/DZ for the low-level data, but instead utilizes MP2/TZ for the high-level data. In this test case, however, the reduced ANI-1x training dataset does not contain forces for MP2/TZ. Therefore, it is not possible to benchmark the MF-DZ/EF-MP2/E MLIP against an SF-MLIP trained directly to MP2/TZ forces (SF-MP2/EF). The benefit of this second test case is that the GDB10to13 test set contains MP2/TZ forces for comparison.

The third test case of MFL utilizes DFT/DZ for low-level data and CCSD(T)*/CBS for high-level data. CC-level forces are not available for either the ANI-1ccx training dataset or the GDB10to13 test dataset. Thus, not only are we not able to train a SF-MLIP directly to CC-level forces for comparison (SF-CC/EF), but we also cannot test the quality of the MLIP predictions against CC-level forces. Instead, we compare the SF-MLIP and MF-MLIP performance for predicting CC-level conformer energy differences for the GDB10to13 dataset.

\subsection{Test case one: DFT/DZ and DFT/TZ}

We begin by investigating MFL with DFT/DZ and DFT/TZ data. Because forces at both levels of fidelity are available for training, this test case allows for direct validation of MFL performance. 

Table~\ref{tbl:DFT/DZ and DFT/TZ MF-TZ} reports the energy and force RMSEs for an SF-MLIP trained just to DFT/TZ energies (SF-TZ/E), an SF-MLIP trained to DFT/TZ energies and forces (SF-TZ/EF), an MF-MLIP trained to DFT/DZ energies and DFT/TZ energies (MF-DZ/E-TZ/E), an MF-MLIP trained to DFT/DZ energies and forces and DFT/TZ energies (MF-DZ/EF-TZ/E), and an MF-MLIP trained to DFT/DZ energies and forces and DFT/TZ energies and forces (MF-DZ/EF-TZ/EF). See Supplementary Material for additional validation results relevant to the DFT/DZ and DFT/TZ test case.

\begin{table}[htb!]
  \caption{Comparison of energy and force root-mean-square errors (RMSEs) of DFT/TZ prediction, using single-fidelity (SF) and multi-fidelity (MF) MLIPs trained with DFT/TZ forces and/or energies with/without DFT/DZ forces and/or energies. MFL with low-level (DFT/DZ) forces improved the prediction on the high-level (DFT/TZ) energies and forces compared to SF training to high-level energies only. RMSEs are computed relative to the DFT/TZ energies and forces for the in-sample test set. Error bars represent 95\% confidence interval from an ensemble of eight models.}
  \label{tbl:DFT/DZ and DFT/TZ MF-TZ}
  \begin{tabular}{c c c }
  \hline
     & Energy-RMSE (kcal/mol) & Force-RMSE (kcal/mol/\AA) \\
  \hline
SF-TZ/E& 2.75 $\pm$ 0.02 & 13.52 $\pm$ 0.50 \\ 
SF-TZ/EF& 1.60 $\pm$ 0.02 & 3.52 $\pm$ 0.09 \\
MF-DZ/E-TZ/E& 2.78 $\pm$ 0.04& 11.92 $\pm$ 0.15\\
MF-DZ/EF-TZ/E& 1.49 $\pm$ 0.04 & 4.16 $\pm$ 0.15 \\
MF-DZ/EF-TZ/EF& 1.34 $\pm$ 0.02 & 3.42 $\pm$ 0.11\\
\hline
  \end{tabular}
\end{table}

A key observation from Table~\ref{tbl:DFT/DZ and DFT/TZ MF-TZ} is that MFL with low-level forces (MF-DZ/EF-TZ/E) clearly outperforms SF training to only high-level energies (SF-TZ/E). Specifically, the energy and force RMSEs for MF-DZ/EF-TZ/E (1.49 kcal/mol and 4.16 kcal/mol/\AA) are approximately a factor of two or three times smaller than for SF-TZ/E (2.75 kcal/mol and 13.52 kcal/mol/\AA), respectively. These results provide strong evidence that, in the common scenario where high-level forces are not available for training, MFL with low-level forces is far superior to SF training to only high-level energies.

Furthermore, MFL with low-level forces achieves similar accuracy as training directly to high-level forces. Specifically, the MF-DZ/EF-TZ/E force errors (4.16 kcal/mol/\AA) are only 18\% higher than the SF-TZ/EF force errors (3.52 kcal/mol/\AA) and the MF-DZ/EF-TZ/E energy errors (1.49 kcal/mol) are actually 7\% lower than the SF-TZ/EF energy errors (1.60 kcal/mol). This performance is quite remarkable considering that MF-DZ/EF-TZ/E was not trained with DFT/TZ forces. Therefore, in the absence of high-level forces, MFL with low-level forces is sufficient for training an MLIP that predicts accurate high-level energies and forces.

By contrast, MFL with only energies at both levels of theory (MF-DZ/E-TZ/E) provides limited improvement compared to SF training without forces (SF-TZ/E). The force RMSE for MF-DZ/E-TZ/E remains too large (approximately 12 kcal/mol/\AA) for practical use. The lack of improvement for MFL without low-level forces is consistent with previous work.~\cite{jacobson_leveraging_2023}. Thus, alternative training methods should be investigated if forces are not available at either level of theory.

Although the performance of MFL with low-level forces is impressive, SF training directly to high-level forces is still marginally better for force prediction. If high-level forces are available for training, however, it is also possible to perform MFL with both low-level and high-level forces (MF-DZ/EF-TZ/EF). In fact, the MF-DZ/EF-TZ/EF errors are slightly lower than the SF-TZ/EF errors for both energy (1.34 kcal/mol vs 1.60 kcal/mol) and forces (3.42 kcal/mol/\AA~vs 3.52 kcal/mol/\AA). Therefore, MFL with both low-level and high-level forces outperforms SF learning to each level of theory separately. 

\subsection{Test case two: DFT/DZ and MP2/TZ}

We now investigate MFL with DFT/DZ and MP2/TZ data. Although the training dataset does not contain any MP2/TZ force data, this test case allows for validation against the MP2/TZ forces in the GDB10to13 dataset. In the first test case, the two levels of theory (DFT/DZ and DFT/TZ) are relatively similar and computationally inexpensive. In this test case, however, the MP2/TZ forces are quite different from the DFT/DZ forces (RMSE of 5.9 kcal/mol/\AA~when comparing DFT/DZ and MP2/TZ forces in GDB10to13). The more expensive MP2/TZ forces are also assumed to be more accurate than DFT/TZ, considering MP2 is a higher level of theory and both methods use similar triple zeta basis sets. Therefore, this test case presents a more interesting and practical challenge for MFL than test case one. 

Table~\ref{tbl:DFT/DZ and MP2/TZ} reports the energy and force RMSEs for a single-fidelity MLIP trained just to MP2/TZ energies (SF-MP2/E) and a multi-fidelity MLIP trained to DFT/DZ energies and forces and MP2/TZ energies (MF-DZ/EF-MP2/E). See Supplementary Material for additional validation results relevant to the DFT/DZ and MP2/TZ test case.

\begin{table}[htb!]
  \caption{Comparison of energy and force root-mean-square errors (RMSEs) of MP2/TZ prediction, using single-fidelity (SF) and multi-fidelity (MF) MLIPs trained with MP2/TZ energies with/without DFT/DZ forces and energies. MFL with low-level (DFT/DZ) forces improved the prediction on the high-level (MP2/TZ) energies and forces compared to SF training to high-level energies only. No MP2/TZ forces were used during training of MLIPs. Energy RMSEs are computed relative to the MP2/TZ energies for the in-sample test set. Force RMSEs are computed relative to the MP2/TZ forces in the GDB10to13 out-of-sample test set. Error bars represent 95\% confidence interval from an ensemble of eight models.}
  \label{tbl:DFT/DZ and MP2/TZ}
  \begin{tabular}{c c c }
  \hline
     & Energy-RMSE (kcal/mol) & Force-RMSE (kcal/mol/\AA) \\
  \hline
SF-MP2/E & 2.95 $\pm$ 0.02 & 8.59 $\pm$ 0.42 \\
MF-DZ/EF-MP2/E & 1.72 $\pm$ 0.02 & 3.42 $\pm$ 0.09 \\
\hline
  \end{tabular}
\end{table}

Consistent with the first test case, the results in Table~\ref{tbl:DFT/DZ and MP2/TZ} demonstrate that MFL with low-level (DFT/DZ) forces is significantly more accurate than SF training to only high-level (MP2/TZ) energies. Specifically, both the energy and force RMSEs for MF-DZ/EF-MP2/E (1.72 kcal/mol and 3.42 kcal/mol/\AA) are approximately a factor of two lower than for SF-MP2/E (2.95 kcal/mol and 8.59 kcal/mol/\AA), respectively. Furthermore, the force RMSE for MF-DZ/EF-MP2/E on the GDB10to13 dataset (3.42 kcal/mol/\AA) is significantly lower than the RMSE between DFT/DZ forces and MP2/TZ forces (5.9 kcal/mol/\AA). Thus, MFL is outperforming the accuracy that could be achieved by simply learning the low-level (DFT/DZ) forces. 

\subsection{Test case three: DFT/DZ and CCSD(T)*/CBS}

Having verified that MFL quantifiably improves the high-level energies and forces for the first two test cases, we now aim at the most challenging and important test case, namely, MFL with DFT/DZ and gold-standard CCSD(T)*/CBS data. Because CC is prohibitively expensive, no CC-level force data are available for direct validation in either the ANI-1ccx training dataset or the GDB10to13 test dataset. However, Supplementary Material provides indirect evidence that MFL with low-level forces improves CC-level forces.

We perform two additional tests using CCSD(T)*/CBS energies for conformers from the GDB10to13 dataset and torsion scans from the Sellers et al. dataset.~\cite{sellers_comparison_2017} The conformer energies are reported as energy differences $(\Delta E_{\rm conf})$ between all possible pairs of conformers. The torsion energies are reported as energy differences $(\Delta E_{\rm tors})$ relative to the minimum energy for a given torsion scan. Predicting accurate conformer and torsion energies is essential for determining the most stable structure of a molecule. Conformer geometry searches can be quite expensive when performed with QM methods, whereas classical FFs may not be sufficiently accurate to correctly identify the minimum-energy conformer geometry.~\cite{lahey2020} Thus, an MLIP that can predict accurate CC-level conformer and torsion energy differences would be extremely valuable for real-life applications, such as drug discovery.~\cite{rai2022,mann_egret-1_2025}

Table~\ref{tbl:DFT/DZ and CC} reports the energy and $\Delta E_{\rm conf}$ RMSEs for a single-fidelity MLIP trained to CCSD(T)*/CBS energies (SF-CC/E), a multi-fidelity MLIP trained to DFT/DZ energies and CCSD(T)*/CBS energies (MF-DZ/E-CC/E), and a multi-fidelity MLIP trained to DFT/DZ energies and forces and CCSD(T)*/CBS energies (MF-DZ/EF-CC/E). See Supplementary Material for additional validation results relevant to the DFT/DZ and CCSD(T)*/CBS test case.

\begin{table}[htb!]
  \caption{Comparison of energy, conformer energy differences ($\Delta E_{\rm conf}$), and torsion energy differences ($\Delta E_{\rm tors}$) root-mean-square errors (RMSEs) of CCSD(T)*/CBS prediction, using single-fidelity (SF) and multi-fidelity (MF) MLIPs trained with CCSD(T)*/CBS energies with/without DFT/DZ forces and/or energies. MFL with low-level (DFT/DZ) forces improved the prediction on the high-level (CCSD(T)*/CBS) energies compared to SF training to high-level energies only. No CCSD(T)*/CBS forces were used during training of SF or MF MLIPs. Energy RMSEs are computed relative to the CCSD(T)*/CBS energies for the in-sample test set. $\Delta E_{\rm conf}$ and $\Delta E_{\rm tors}$ RMSEs are computed relative to the CCSD(T)*/CBS energy differences for the GDB10to13 conformer and torsion scan out-of-sample test sets, respectively. Error bars represent 95\% confidence interval from an ensemble of eight models.}
  \label{tbl:DFT/DZ and CC}
  \begin{tabular}{c c c c }
  \hline
     & Energy-RMSE (kcal/mol) & $\Delta E_{\rm conf}$-RMSE (kcal/mol) & $\Delta E_{\rm tors}$-RMSE \\
  \hline
SF-CC/E & 2.78 $\pm$ 0.05 & 2.65 $\pm$ 0.04 & 0.71 $\pm$ 0.03\\
MF-DZ/E-CC/E & 2.76 $\pm$ 0.03 & 2.60 $\pm$ 0.03 & 0.67 $\pm$ 0.04 \\
MF-DZ/EF-CC/E & 1.52 $\pm$ 0.03 & 1.60 $\pm$ 0.03 & 0.40 $\pm$ 0.02 \\
\hline
  \end{tabular}
\end{table}

Similar to the results for the previous two test cases, the results in Table~\ref{tbl:DFT/DZ and CC} demonstrate that MFL with low-level (DFT/DZ) forces is significantly more accurate than SF training to only high-level (CCSD(T)*/CBS) energies. Specifically, the MF-DZ/EF-CC/E errors are considerably lower than the SF-CC/E errors on CCSD(T)*/CBS energies (1.52 kcal/mol vs 2.78 kcal/mol), CCSD(T)*/CBS conformer energy differences (1.60 kcal/mol vs 2.65 kcal/mol), and CCSD(T)*/CBS torsion scan energy differences (0.40 kcal/mol vs 0.71 kcal/mol), respectively. The MF-DZ/EF-CC/E energy, $\Delta E_{\rm conf}$, and $\Delta E_{\rm tors}$ RMSEs (1.52 kcal/mol, 1.60 kcal/mol, and 0.40 kcal/mol) are also substantially lower than the RMSEs for the ANI-1ccx MLIP (2.57 kcal/mol, 2.62 kcal/mol, and 0.77 kcal/mol), respectively. See Supplementary Material for additional comparisons between HIP-NN and ANI models. By contrast, MFL without forces (i.e., only energies for both levels of theory) is essentially as accurate as SF training to just the high-level energies, consistent with previous work.~\cite{jacobson_leveraging_2023} This result demonstrates that including low-level forces is essential for MFL to achieve significant improvement in accuracy.


\section{Discussion} 

\begin{figure}[htb!]
    \includegraphics[width=3in]{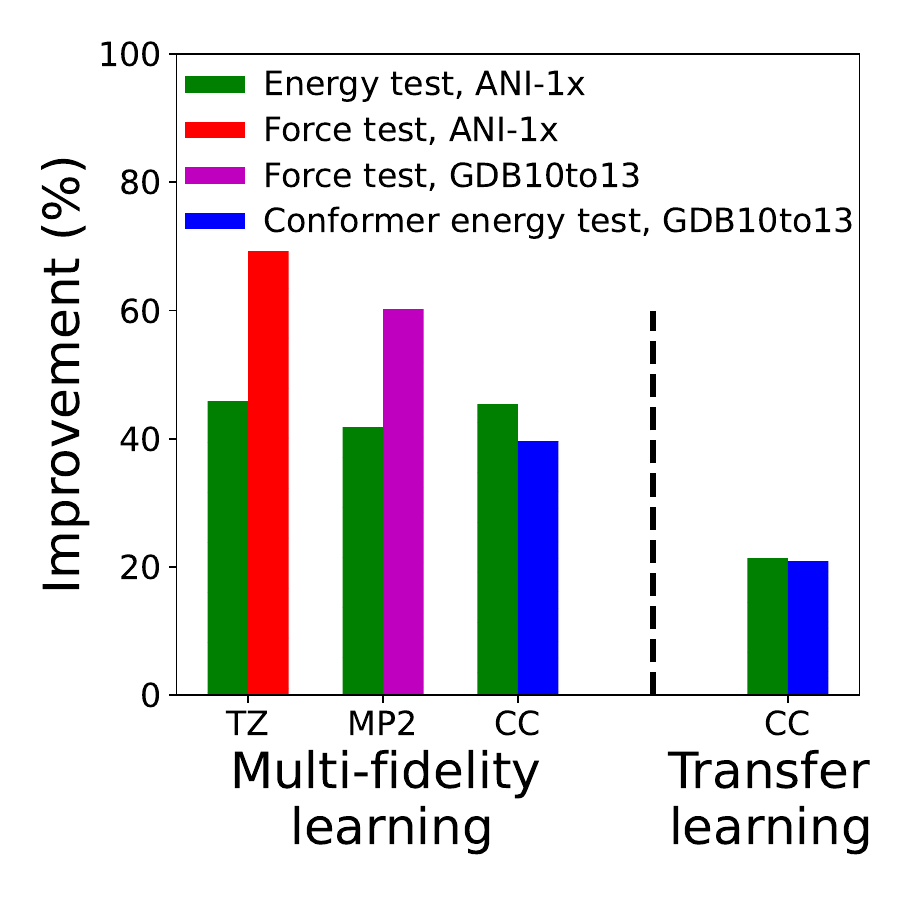}
    \caption{Comparison of percent improvement for multi-fidelity learning (MFL) with three test cases and for transfer learning (TL). Percent improvements are computed for the energy, forces, and conformer energy differences $(\Delta E_{\rm conf})$. Force improvement is computed for both the in-sample test dataset (held-out data from ANI-1x) and for the out-of-sample GDB10to13 test dataset. Labels on horizontal axis represent the higher level of theory used in training, where TZ = DFT/TZ, MP2 = MP2/TZ, and CC = CCSD(T)*/CBS. Improvement percentage for MFL is calculated by $($RMSE$_{\rm SF}-$RMSE$_{\rm MF})/$RMSE$_{\rm SF} \times 100\%$, where ``SF'' corresponds to SF-X/E and ``MF'' corresponds to MF-DZ/EF-X/E, where `X' represents either TZ, MP2, or CC. The improvement percentage for TL is calculated by $($RMSE$_{\rm no TL}-$RMSE$_{\rm TL})/$RMSE$_{\rm no TL} \times 100\%$, where ``TL'' and ``noTL'' correspond to the ANI-1ccx and ANI-1ccx-R MLIPs, respectively~\cite{smith2019approaching}. RMSEs for ANI-1x tests are computed with the corresponding TZ, MP2, or CC energies and the DFT/TZ forces. RMSEs for GDB10to13 tests are computed with the MP2/TZ forces and CCSD(T)*/CBS conformer energy differences. MFL shows strong improvements in accuracy (45\%-70\%) when trained to low-level (DFT/DZ) energies and forces as well as energies from a second higher level of theory (TZ, MP2, or CC).}
    \label{fig:Improvement factor rough chart}
\end{figure}

These three test cases clearly demonstrate that MFL with low-level forces and high-level energies outperforms SF training with only high-level energies. Moreover, MFL with low-level forces is of similar accuracy as SF training directly to high-level forces, in cases where those are available for comparison. To help quantify the improvement MFL provides in all three test cases, Figure~\ref{fig:Improvement factor rough chart} presents the percent decrease in RMSE for energies, forces, and conformer energy differences between each MF-MLIP trained with low-level (DFT/DZ) forces and the corresponding SF-MLIP trained with only high-level (either DFT/TZ, MP2/TZ, or CCSD(T)*/CBS) energies. See Supplementary Material for additional analysis of percent improvement. 

The percent improvement in energy and conformer energy differences with MFL is between 40\% and 45\% for all three test cases. The percent improvement in forces with MFL is 70\% for MF-DZ/EF-TZ/E tested on the ANI-1x DFT/TZ forces and 60\% for MF-DZ/EF-MP2/E tested on the GDB10to13 MP2/TZ forces. For comparison, the percent improvement with TL is more than a factor of two smaller than with any of the MFL test cases. Specifically, based on the reported errors for the ANI-1ccx and ANI-1ccx-R MLIPs~\cite{smith2019approaching}, which were trained with and without TL, respectively, the percent improvement with TL is between 15-20\% for energies and conformer energies. For a truly fair comparison between TL and MFL, however, future work could investigate the percent improvement for TL that is achieved when pre-training with the same HIP-NN architecture and the same reduced ANI-1x dataset with both DFT/DZ energies and DFT/DZ forces. 

One unexpected result is that MFL leads to a measurable decrease in accuracy on the low-level forces themselves (see Supplementary Material). Specifically, the RMSEs on the DFT/DZ forces for MF-DZ/EF-TZ/E (3.95 kcal/mol/\AA), MF-DZ/EF-MP2/E (3.91 kcal/mol/\AA), and MF-DZ/EF-CC/E (3.76 kcal/mol/\AA) are all markedly higher than the force RMSE for SF-DZ/EF (3.17 kcal/mol/\AA). This increase in error on the low-level (DFT/DZ) forces suggests that the current MFL approach requires a compromise in accuracy on the low-level forces to achieve improved accuracy on the high-level forces. Future work could attempt to elucidate whether this is a fundamental limitation of MFL, a limitation of our implementation, or a limitation in the model architecture. Some possible considerations to help reduce the low-level force errors are to use a larger HIP-NN architecture (i.e., more interactions layers, more atom layers, more features), and/or modify the number of shared parameters, and/or include non-linear layers for the fidelity-specific output nodes. However, we emphasize that the primary goal of this work is to improve the high-level energy and force predictions. Indeed, if only accurate prediction of the low-level energy and forces is the desired outcome, SF learning would suffice.

For completeness, we also investigate the potential benefit of applying MFL with forces from more than one level of theory. In this case, the energies and forces for both DFT/DZ and DFT/TZ are included in training along with the CCSD(T)*/CBS energies. Notably, the CC-level errors are actually considerably worse for MFL with three datasets than MFL with just two datasets (see Supplementary Material). With twice as much DFT-level data, the worse performance with three datasets is likely attributed to a data imbalance. An alternative explanation is that the DFT/DZ and DFT/TZ forces are providing contradictory information that MFL simply cannot resolve. Another possibility is that the MLIP has reached its capacity to match the different datasets with the chosen architecture and hyperparameters. However, with the current MFL approach, low-level forces from just a single level of theory are sufficient to achieve the desired accuracy on the high-level forces.

\section{Conclusions and Future Work}

In the age where datasets with low-level quantum mechanical forces are in abundance but high-level forces are scarce, multi-fidelity learning is a powerful technique to develop machine learning interatomic potentials with gold-standard coupled-cluster-level accuracy. Although previous studies with multi-fidelity learning to only energy data did not report significant improvements compared to single-fidelity learning, the current work shows that the inclusion of low-level forces is all that is needed to achieve high-level accuracy when the two datasets cover the same configuration space. 

The results of all three test cases are complementary, and the conclusions are unanimous. First, multi-fidelity learning with low-level forces achieves a factor of two improvement in high-level energies and forces compared to single-fidelity learning with only high-level energies. Second, multi-fidelity learning with low-level forces achieves similar accuracy as single-fidelity learning directly to high-level forces. Third, multi-fidelity learning with both low-level and high-level forces achieves the highest accuracy. Fourth, multi-fidelity learning without forces does not provide significant benefit compared to single-fidelity learning with only energies.

While the present study focused on multi-fidelity learning with just two or three quantum mechanical datasets, multi-fidelity learning is capable of leveraging several datasets with vastly different amounts of data. Future work could investigate alternative multi-fidelity learning training protocols that better account for data volume imbalance between different levels of theory. This study considered multi-fidelity learning when both the low-level and high-level data exist for the same molecular structures. While this approach clearly improves accuracy on the types of systems included in the training set, it is unlikely to improve transferability. Future work could consider how multi-fidelity learning can leverage disparate datasets to share knowledge from different regions of chemical space. Ideally, the molecular structures would cover completely different portions of the potential energy surface, although some overlap may be necessary for joint learning.~\cite{shoghi_molecules_2023} For example, future work could determine if multi-fidelity learning with low-level forces and high-level energies is beneficial when training to both non-reactive (e.g., ANI-1ccx) and reactive (e.g., Transition1x)~\cite{schreiner2022transition1x} datasets or when training to both \textit{in vacuo} (e.g., ANI-1ccx) and periodic boundary (e.g., high-pressure MD)~\cite{pilsun_neural_2021,hamilton2023,willman_machine_2022} datasets. Future work could also test the limits of multi-fidelity learning by using low-level forces that are computed with extremely inexpensive but even more approximate methods, such as classical force fields or semi-empirical quantum chemistry.~\cite{thiel2014} Future work could also consider including in the multi-fidelity dataset so-called ``atomic energies.'' Although atomic energies are not physical quantities that can be computed with quantum mechanical methods, previous studies have shown the advantage of (pre-)training to atomic energies~\cite{gardner_synthetic_2023,gardner_synthetic_2024} computed with either a classical force field~\cite{jung2025} or a ``teacher'' machine learning interatomic potential.~\cite{matin_ensemble_2025,luan2025} Furthermore, in light of the recent advances training with Hessian data,~\cite{rodriguez_does_2025,amin_towards_2025} future work could also investigate multi-fidelity learning with low-level Hessians.

\clearpage
\newpage

\section*{Data availability}

The ANI-1x and GDB10to13 datasets are publicly available at, respectively, \url{https://springernature.figshare.com/articles/dataset/ANI-1x_Dataset_Release/10047041} and \url{https://github.com/isayev/COMP6/tree/master/COMP6v1/GDB10to13}. 

The data that support the findings of this study will be openly available following an embargo at the following URL/DOI: \url{https://github.com/lanl/hippynn/examples/multifidelity_ani1x_training.py}.

\section*{Acknowledgments}

M.M., S.M., A.E.A.A., B.N., K.B., N.L., and R.A.M. acknowledge support from the US Department of Energy, Office of Science, Basic Energy Sciences, Chemical Sciences, Geosciences, and Biosciences Division under Triad National Security, LLC (‘Triad’) contract grant 89233218CNA000001 (FWP: LANLE3F2 and LANLE8AN). M.M. gratefully acknowledges the resources of the Los Alamos National Laboratory (LANL) Computational Science summer student program.  The work at LANL was supported by the LANL Laboratory Directed Research and Development (LDRD) Project 20230290ER. Work at LANL was performed in part at the Center for Nonlinear Studies and the Center for Integrated Nanotechnologies, a US Department of Energy Office of Science user facility at LANL. This research used resources provided by the Darwin testbed at LANL which is funded by the Computational Systems and Software Environments subprogram of LANL's Advanced Simulation and Computing program (NNSA/DOE). This research used resources of the Oak Ridge Leadership Computing Facility at the Oak Ridge National Laboratory, which is supported by the Office of Science of the U.S. Department of Energy under Contract No. DE-AC05-00OR22725.

\clearpage
\newpage

\section*{Appendix}
\label{sec:appendix}

\subsection*{Terminology}

Although the terms multi-fidelity, multi-level, multi-task, and multi-head are used somewhat interchangeably in the literature, we briefly explain some subtle differences between these methods within the context of MLIP development. 

Typical MLIPs predict only a single energy. To ensure conservation of energy, forces are not, or at least ``should'' not be, predicted directly as a second output. Rather, the MLIP forces are obtained by differentiation of the energy with respect to the atomic positions. Thus, most MLIPs possess a single output node, often referred to as a ``head'' in the context of neural networks. Although the loss function might include multiple error terms that depend on energies and forces (and sometimes stresses), these error terms do not require separate training ``tasks''. Thus, traditional training is neither multi-task nor multi-head learning, although it may still be considered a multi-objective optimization. 

Multi-head learning is performed whenever an MLIP directly predicts multiple outputs. These multiple output heads can predict the energy for different QM levels of theory, as in the case of multi-fidelity (or multi-level) learning. However, in addition to predicting energies, these output heads can also predict atomic or molecular properties (e.g., dipole moment, electronegativity).~\cite{tang_approaching_2024,Unke2019PhysNet:Charges,zubatiuk_development_2021} As learning these multiple properties requires separate training ``tasks'', this example of multi-head learning is also referred to as multi-task learning. However, as each property is typically computed using a single QM level of theory, this is not an example of multi-fidelity learning. Training to energies and forces at multiple levels of theory also requires separate training ``tasks.'' Therefore, although multi-head learning and multi-task learning are not necessarily the same as multi-fidelity learning, the multi-fidelity learning approach used in this study can also be considered multi-level, multi-head, and multi-task learning. 

Not all multi-fidelity learning approaches utilize multi-head learning. For example, an alternative multi-fidelity approach is to provide the desired level of theory directly to the neural network as an input, such that a single output head predicts the energy corresponding to the specified level of theory.~\cite{chen_one_2025} 

\clearpage
\newpage

\subsection*{HIP-NN Overview}

All models in this study utilize the ``Hierarchically Interacting Particle Neural Network with Tensor Sensitivity Information'' (HIP-NN-TS) architecture,~\cite{chigaev_lightweight_2023} an extension of the original HIP-NN architecture.~\cite{lubbers2018hierarchical} The original HIP-NN architecture is limited in terms of how information about atomic neighborhoods is collected by the model. The extended HIP-NN-TS architecture can incorporate either scalar, dipole, or quadrupole information about the feature distribution in the environment of an atom. HIP-NN-TS with tensor sensitivity of zero (i.e., scalar information) is equivalent to the original HIP-NN model. The accuracy (and computational cost) of HIP-NN-TS increases somewhat with increasing tensor sensitivity. To maximize accuracy, all HIP-NN-TS models reported in the main text utilize tensor sensitivity of $\ell=2$ (that is, using scalar, dipole, and quadrupole information). For comparison, Supplementary Material provides results for other HIP-NN-TS models with tensor sensitivity of zero (scalar information) and one (scalar and dipole information). For brevity, we refer to HIP-NN-TS simply as HIP-NN throughout the main text. 

\begin{figure}[htb!]
    \includegraphics[width=4in]{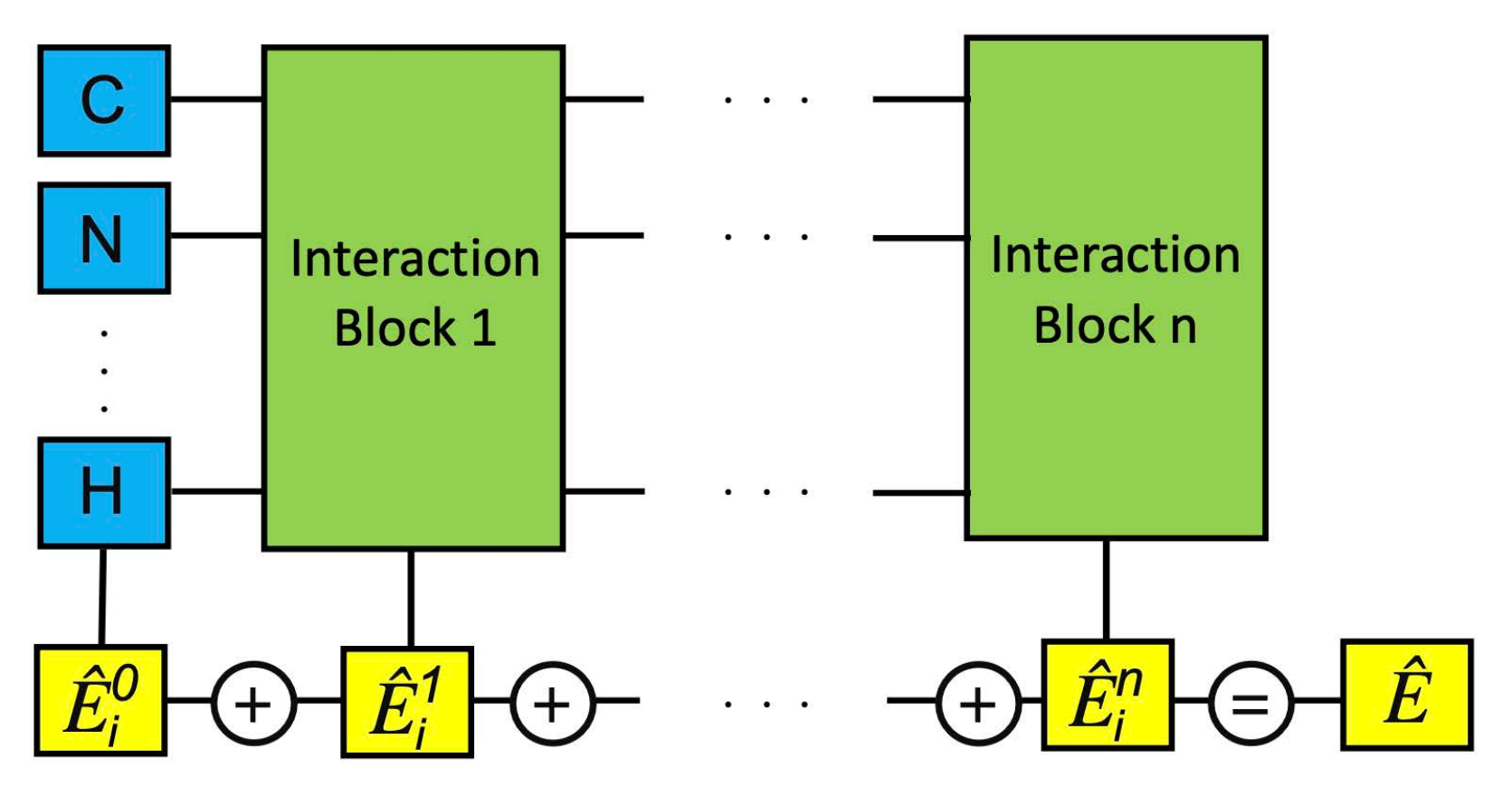}
    \includegraphics[width=2in]{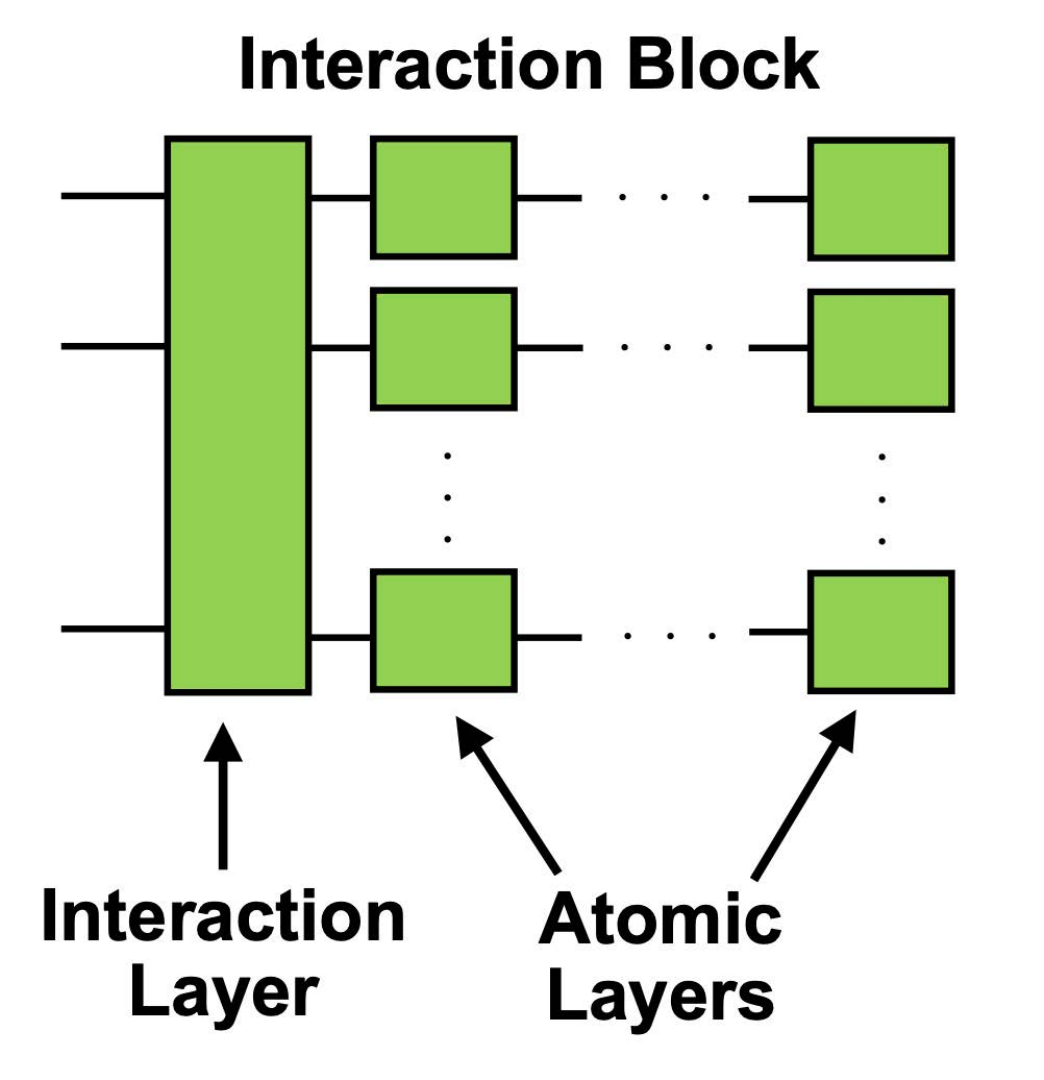}
    \caption{(Left) Standard HIP-NN single-fidelity architecture with interaction blocks 1 to n. (Right) Interaction block containing one interaction layer followed by $n_{\rm layers}$ atomic layers.}
    \label{fig:General HIPNN architecture}
\end{figure}

HIP-NN is a message-passing graph-based neural network that consists of different types of layers. At each layer, HIP-NN computes an atomic feature vector $z_i$ for each atom $i$. This atomic feature vector is an abstract numeric representation of the local chemical environment around atom $i$. $z_i$ is composed of $n_{\rm features}$ number of scalar components, $z_{i,a}$, for $a=1...n_{\rm features}$. The one exception is the input layer (see blue squares in Figure \ref{fig:General HIPNN architecture}). The input feature vector encodes the atomic species (H, C, N, and O in this work) of each atom $i$ using one-hot encoding into a feature dimension indexed by $a$. Thus, the dimensionality of the initial feature vector is equal to the number of element types (four in this work).

A key feature of HIP-NN is the interaction block (see green rectangles in left panel of Figure \ref{fig:General HIPNN architecture}). Interaction blocks contain one interaction layer followed by $n_{\rm layers}$ number of atomic layers (see right panel in Figure \ref{fig:General HIPNN architecture}).

Interaction layers enable message passing between atoms within a predefined cut-off distance (6.5 \AA~in this work), mixing the feature vectors of neighboring atoms. Each interaction layer transforms the atomic feature vector, $z_i$, into an updated atomic feature vector, $z_i'$, which is then passed to the next atomic layer. This transformation is given by 
\begin{equation}
\label{int_layer}
    z'_{i,a} = f\left(I_{i,a}(z,r)+\sum_b W_{ab}z_{i,b} + B_a\right)
\end{equation}
where $W$ is a trainable weight matrix, $B$ is a trainable bias vector, and $f$ is a non-linear activation function. Both $W$ and $B$ operate locally for each atom $i$. The interaction term, $I$, provides the important information about the local atomic environment by incorporating contributions from neighboring atoms within the cut-off distance. The interaction term is defined as
\begin{equation}
    I_{i,a}(z,r) = \sum_{j,b} v_{ab}(r_{ij})z_{j,b}.
\end{equation}
Here, the interaction term aggregates information from the feature vectors $z_j$ of neighboring atoms $j$ (where  $j\neq i$) within the cut-off distance. The function $v_{ab}$ depends on the interatomic distance $r_{ij} = |r_i-r_j|$, and is expanded as
\begin{equation}
    v_{ab}(r_{ij}) = \sum_\nu V_{ab}^\nu s^\nu (r_{ij}),
\end{equation}
where $V^\nu$ is trainable weight matrix and $s^\nu(r_{ij})$ are a finite basis of trainable sensitivity functions.

Similar to interaction layers, atomic layers also apply a local transformation to the feature vector of each atom, $z_i$ to $z_i'$. This transformation is given by: 
\begin{equation}
\label{atomic_layer}
    z'_{i,a} = f\left(\sum_n w_{ab}z_{i,n} + b_a\right)
\end{equation}
where $w$ is a trainable weight matrix, $b$ is a trainable bias vector, and $f$ is a non-linear activation function.

In the spirit of many-body theory, HIP-NN adheres to the assumption that the total energy of a system can be separated into multiple hierarchical energy contributions. To this end, HIP-NN predicts a hierarchical energy for each atom $i$ (yellow boxes in Figure~\ref{fig:General HIPNN architecture}). This hierarchical energy is obtained from a linear regression of the atomic feature vector according to
\begin{equation}
\label{linear_output_layer}
    E^n_i = \sum_a w_a z_{i,a}+b
\end{equation}
where $E^n_i$ is the $n^{\rm th}$ hierarchical energy for atom $i$, $w$ is a trainable weight matrix and $b$ is a trainable bias vector. For $n=0$, $z_{i,a}$ corresponds to the one-hot encoding input feature vector. For $n>0$, $z_{i,a}$ corresponds to the transformed feature vector following the $n^{\rm th}$ interaction block. The hierarchical energies for a given atom $i$ are summed together to give the so-called atomic energy
\begin{equation}
    E_i = \sum_n E^n_i
\end{equation}
where, since all models in this work have only two interaction blocks, the sum is over $n=$ 0, 1, and 2. 
Finally, as is common practice for most MLIPs and classical FFs, these atomic energies are summed together over all atoms to obtain the total system energy
\begin{equation}
    E = \sum_i E_i
\end{equation} 
To ensure conservation of energy, the forces acting on each atom are not predicted directly by HIP-NN. Instead, the forces are computed, using automatic differentiation, as the negative derivative of the total energy with respect to position for a given atom.

\clearpage
\newpage


\printbibliography

\end{document}


\title{Supplementary Material for:\\ Multi-fidelity learning for interatomic potentials: Low-level forces and high-level energies are all you need} 

\author[1,2]{Mitchell Messerly}
\author[1]{Sakib Matin}
\author[1,3,4]{Alice E. A. Allen}
\author[1]{Benjamin Nebgen}
\author[1,3]{Kipton Barros}
\author[5]{Justin S. Smith}
\author[6]{Nicholas Lubbers}
\author[1,7]{Richard Messerly\thanks{Corresponding author: messerlyra@ornl.gov}}

\affil[1]{Theoretical Division, Los Alamos National Laboratory, Los Alamos, NM 87545, USA}
\affil[2]{Department of Mechanical Engineering, Brigham Young University, Provo, UT 84604, USA}
\affil[3]{Center for Nonlinear Studies, Los Alamos National Laboratory, Los Alamos, New Mexico 87545, United States}
\affil[4]{Max Planck Institute for Polymer Research, Ackermannweg 10, 55128 Mainz, Germany}
\affil[5]{Nvidia Corporation, Santa Clara, CA 95051, United States}
\affil[6]{Computer, Computational, and Statistical Sciences Division, Los Alamos National Laboratory, Los Alamos, New Mexico 87545, United States}
\affil[7]{National Center for Computational Sciences Division, Oak Ridge National Laboratory, Oak Ridge, TN 37830, USA}

\date{This manuscript has been authored in part by UT-Battelle, LLC, under contract DE-AC05-00OR22725 with the U.S. Department of Energy (DOE).  The U.S. government retains and the publisher, by accepting the article for publication, acknowledges that the U.S. government retains a nonexclusive, paid-up, irrevocable, worldwide license to publish or reproduce the published form of this manuscript, or allow others to do so, for U.S. government purposes.  DOE will provide public access to these results of federally sponsored research in accordance with the DOE Public Access Plan (http://energy.gov/downloads/doe-public-access-plan).} 
\maketitle

\clearpage
\newpage

\setcounter{figure}{0}
\setcounter{table}{0}
\renewcommand{\thefigure}{S\arabic{figure}}
\renewcommand{\thetable}{S\arabic{table}}

\section{Model and training details} \label{sec:SI-model_training}

Consistent with previous work, all HIP-NN models in this work have 20 sensitivity functions ($n_{\rm sensitivities} = 20)$, 120-dimensional atomic feature vector ($n_{\rm features}=120$), two interaction blocks $(n_{\rm interaction}=2)$, and four atomic layers in each interaction block $(n_{\rm layers} = 4)$.~\cite{chigaev_lightweight_2023} 

Both the SF and MF models contain approximately 575k parameters. For the MF models, the vast majority ($\approx$99\%) of these $\approx$575k parameters are shared parameters. Each fidelity level in the MF models contains 246 fidelity-specific parameters. Thus, for MF models trained with two levels of fidelity, only 492 of the $\approx$575k parameters (less than 1\%) are fidelity-specific parameters. 

All models use a soft minimum cut-off distance of 0.75\AA~, a soft maximum cut-off distance of 5.5\AA~, and a hard maximum cut-off distance of 6.5\AA. All models use a softplus activation function for $f$. 

Training is performed with the Adam optimizer with a constant batch size of 256 and an initial learning rate of 0.001. The learning rate is reduced by a factor of two if a new best model is not found after 20 epochs. Training is terminated after a maximum of 2000 epochs or if a new best model is not found after 40 epochs. The best model is identified by the lowest combined energy RMSE for all levels of theory.

An ensemble of eight separate HIP-NN MLIPs is trained for all models in this work. Training for each ensemble member is initialized with a different random seed. Each ensemble member is trained to a different random subset of the dataset and tested against a different random in-sample held-out dataset. The training, validation, and held-out test datasets are obtained from a random 80/10/10 split of the reduced ANI-1x dataset.

\clearpage
\newpage

\section{Multi-fidelity HIP-NN architectures} \label{sec:SI-HIPNN}

\begin{figure}[htb!]
    \includegraphics[width=3in]{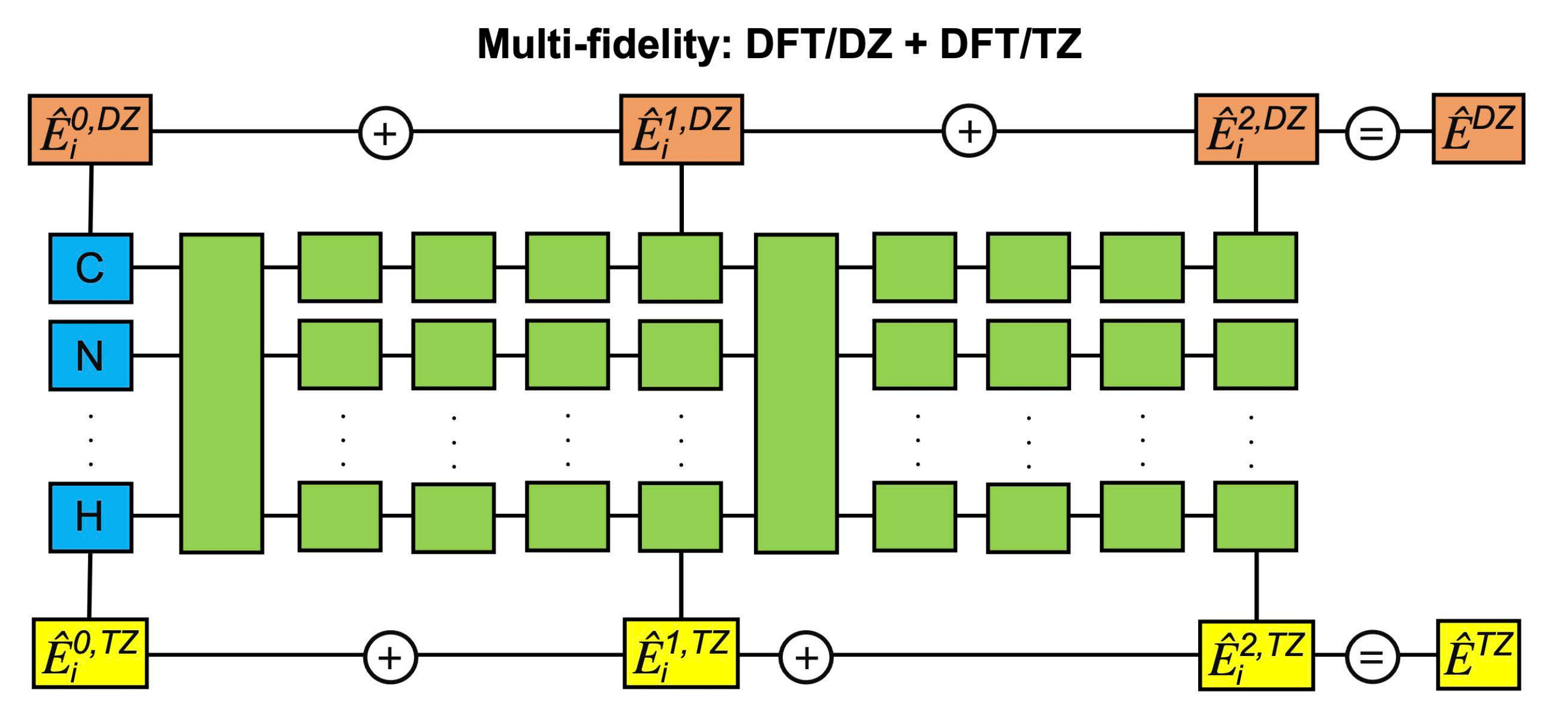}
    \includegraphics[width=3in]{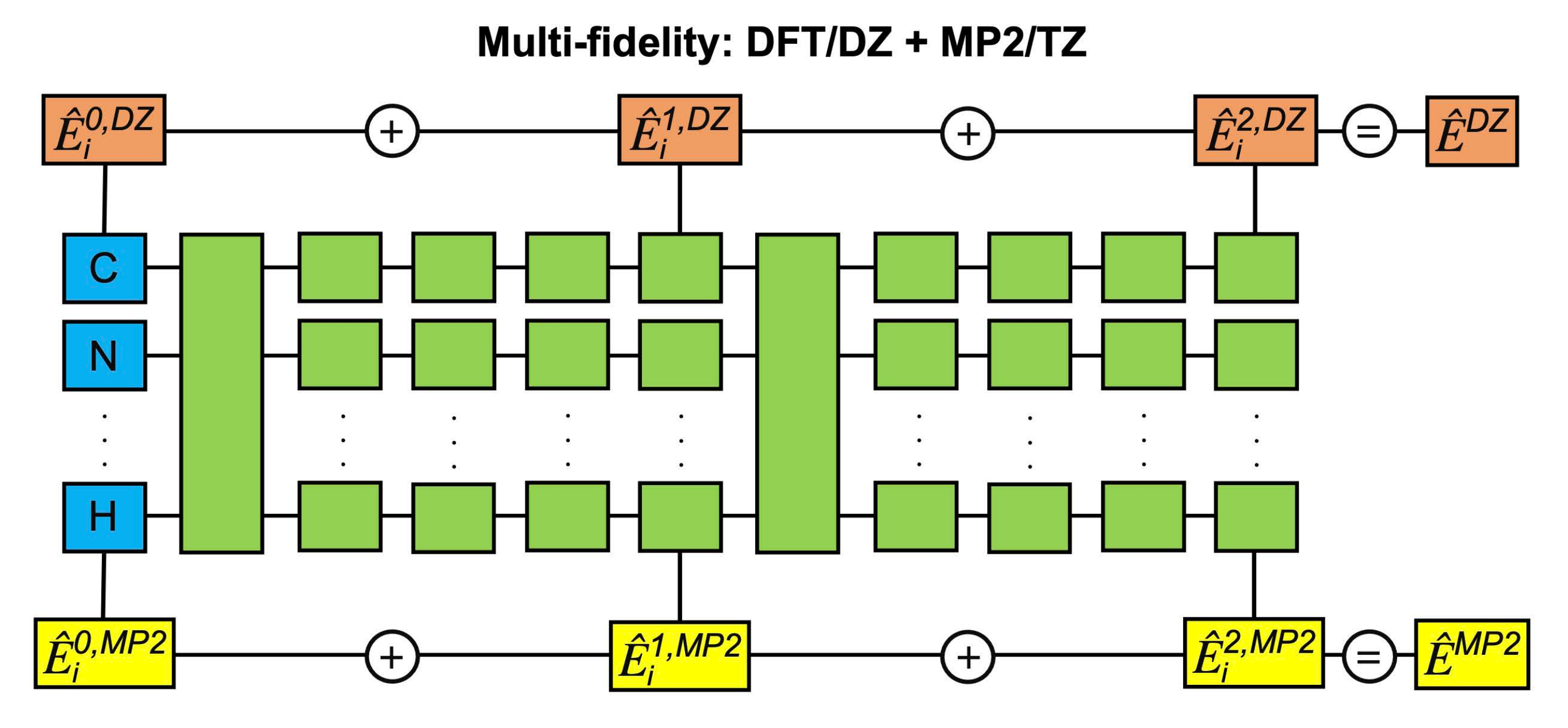}
    \includegraphics[width=3in]{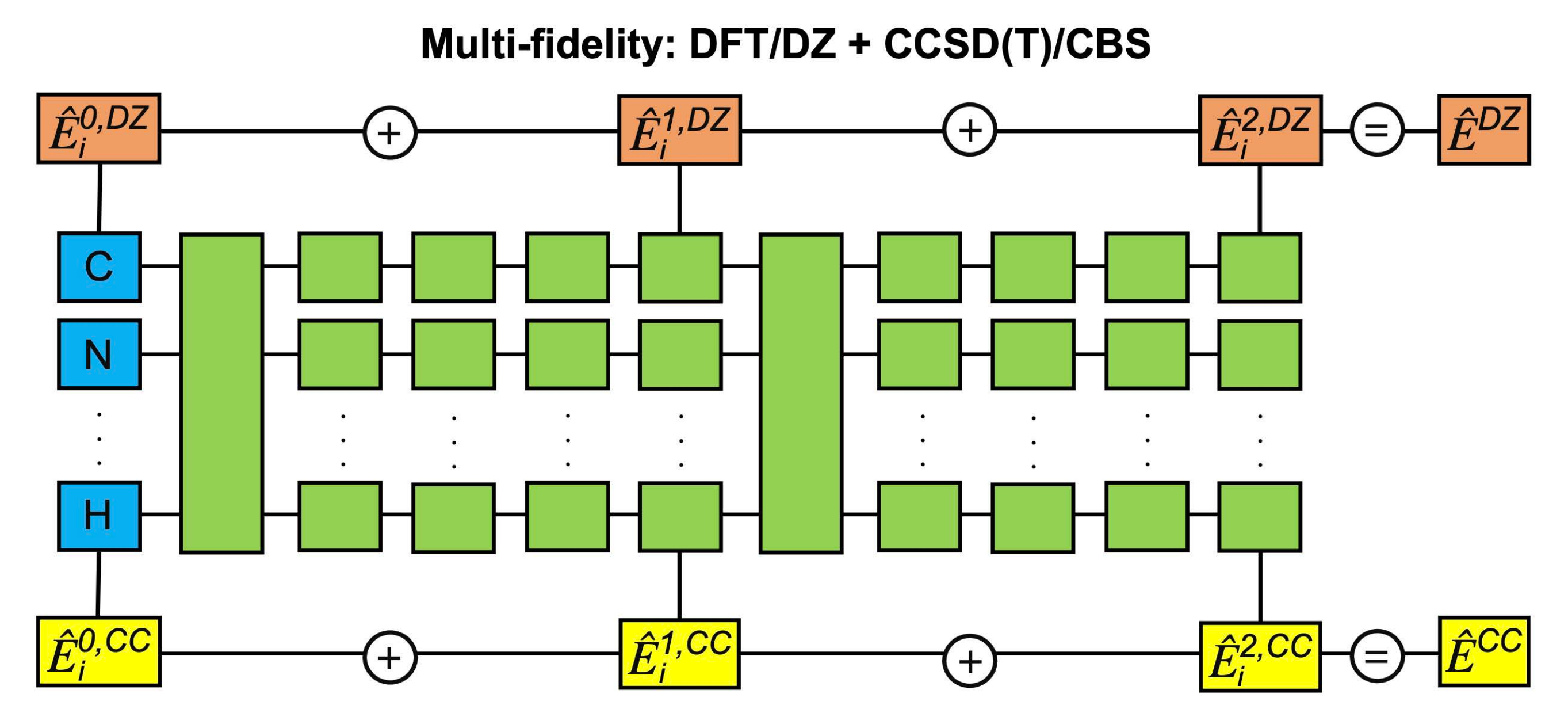}
    \includegraphics[width=3in]{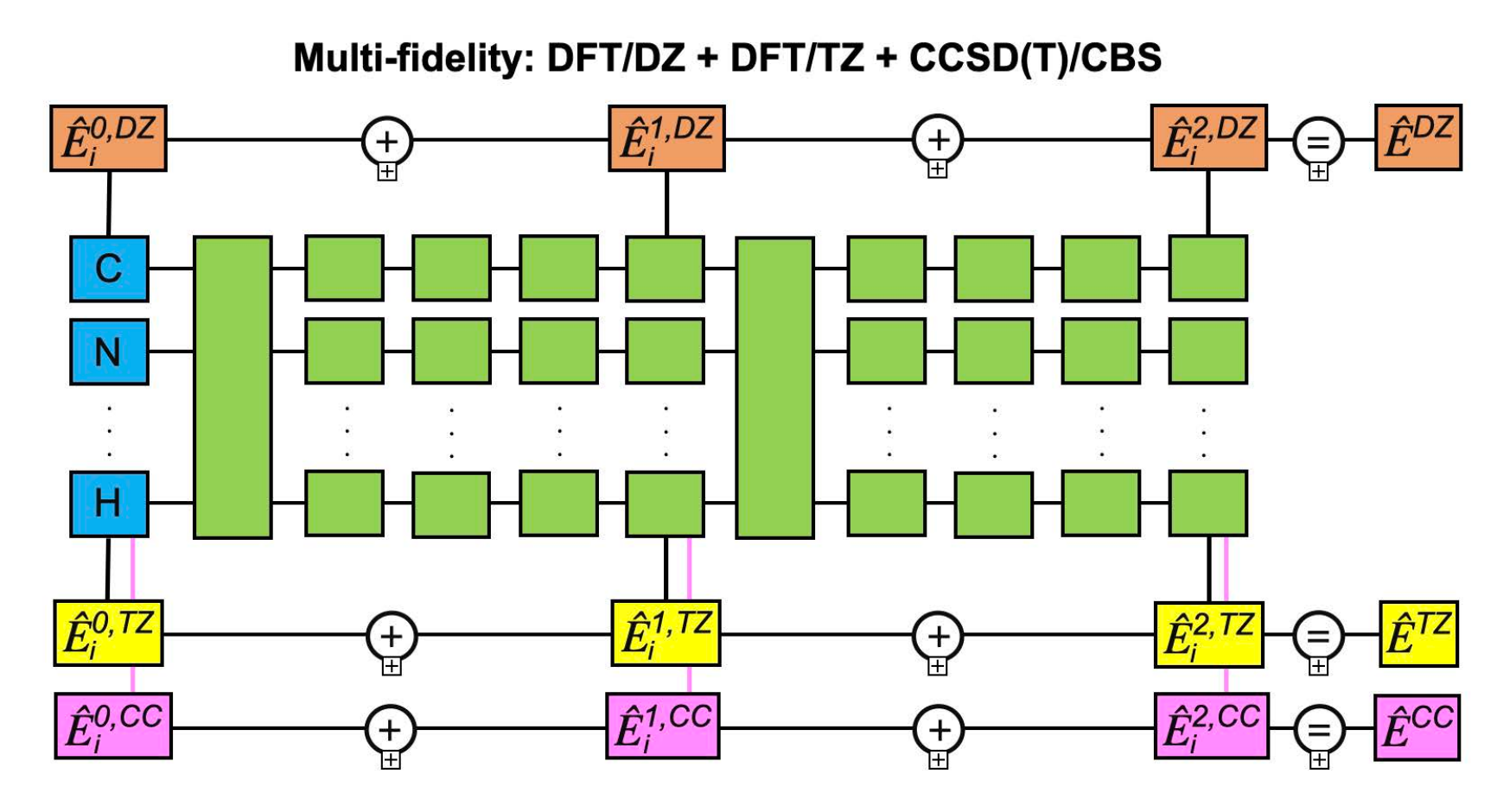}
    \caption{Multi-fidelity HIP-NN architectures for MLIPs trained to (top left) DFT/DZ and DFT/TZ, (top right) DFT/DZ and MP2/TZ, (bottom left) DFT/DZ and CCSD(T)*/CBS, and (bottom right) DFT/DZ, DFT/TZ and CCSD(T)*/CBS. Green boxes denote shared parameters. Brown, yellow, and pink boxes denote fidelity-specific parameters.}
    \label{fig:SI-Specific model architectures}
\end{figure}

\clearpage
\newpage

\section{Further validation of test case one: DFT/DZ and DFT/TZ}

We investigate the data efficiency of MFL by performing a series of MF trainings with different fractions of the reduced ANI-1x dataset ($\approx$460k structures). For the first data-efficiency test, we include the DFT/TZ energy data for all $\approx$460k structures but DFT/DZ energy and force data for only 0\%, 5\%, 10\%, 20\%, 40\%, 80\%, or 100\% of the $\approx$460k structures. For the second data-efficiency test, we include the DFT/DZ energy and force data for all $\approx$460k structures but DFT/TZ energy data for only 5\%, 10\%, 20\%, 40\%, 80\%, or 100\% of the $\approx$460k structures. For both tests, the included structures were selected randomly.  

\begin{figure}[htb!]
    \includegraphics[width=3in]{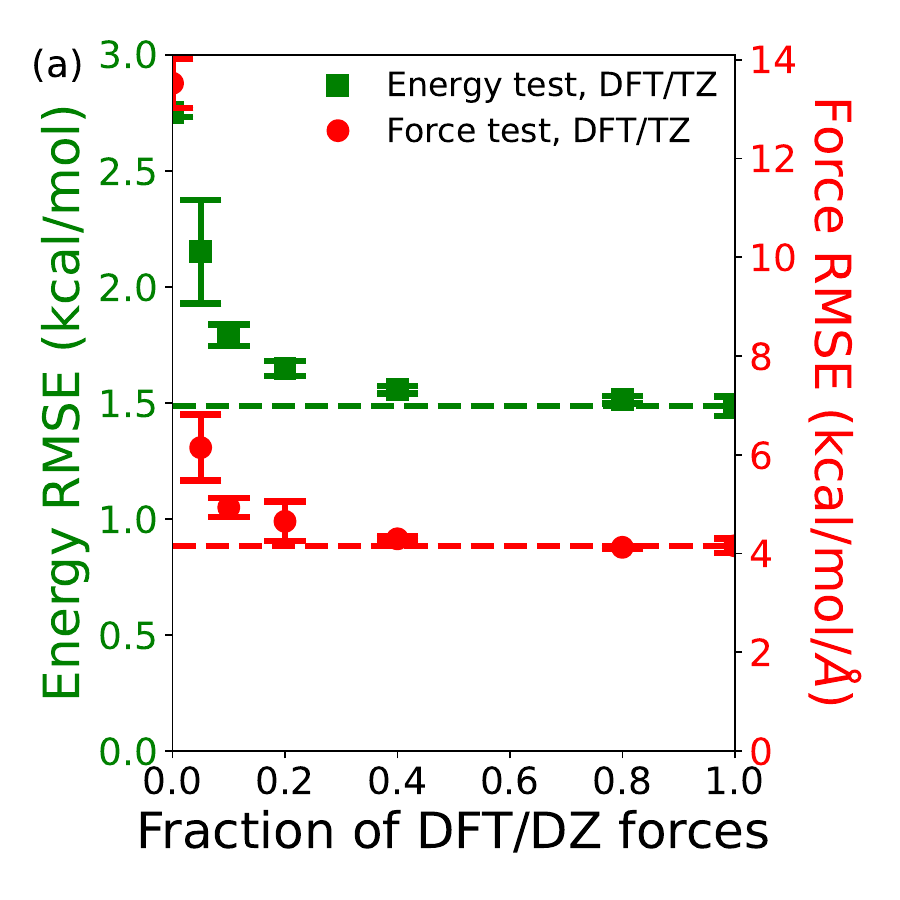}
    \includegraphics[width=3in]{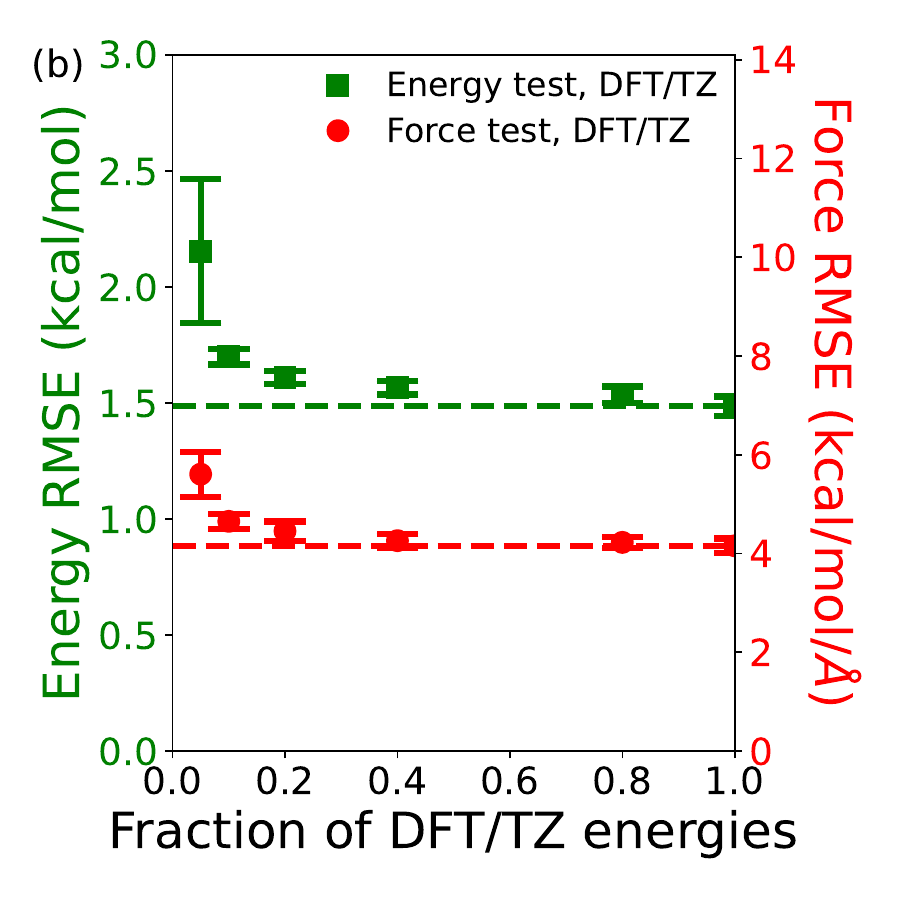}
    \caption{Data-efficiency test for multi-fidelity learning with different fractions of the (a) DFT/DZ forces and (b) DFT/TZ energies. Comparison of energy and force root-mean-square errors (RMSEs) for multi-fidelity MLIPs trained with DFT/DZ energies and forces and DFT/TZ energies. Errors are computed based on the DFT/TZ dataset. Error bars represent 95\% confidence interval from an ensemble of eight models. Dashed lines are the RMSEs reported in the main text for MF-DZ/EF-TZ/E.}
    \label{fig:SI-data_efficiency}
\end{figure}

The results in Figure~\ref{fig:SI-data_efficiency} for both tests are quite similar. Panels (a) and (b) demonstrate that convergence is achieved with $\approx$40\% of the DFT/DZ forces and $\approx$40\% of the DFT/TZ energies, respectively. Thus, the improvement reported in the main text with MFL could be achieved with a factor of two fewer low-level forces or a factor of two fewer high-level energies. Furthermore, if computing high-level energies for all structures is prohibitively expensive, panel (b) suggests that MFL is still extremely beneficial even with only $\approx$10\%-20\% of the high-level energies. Utilizing active learning instead of random selection to determine which structures to include could potentially lead to convergence with even fewer structures. Specifically, only structures with the highest ensemble uncertainty for the high-fidelity energy output node would be labeled with expensive high-level calculations. The results for both tests also demonstrate that complete overlap in configurational space between the low- and high-fidelity data is not required to achieve significant improvement with MFL.

\clearpage
\newpage

To further validate the results for test case one, we also perform the reverse test, using MFL with DFT/TZ forces instead of DFT/DZ forces and test the MLIP on the DFT/DZ data (see Supplementary Material Table~\ref{tbl:DFT/DZ and DFT/TZ MF-DZ}). Consistent with the test case one results in the main text, the energy and force errors for the MF-MLIP trained with DFT/TZ forces (MF-DZ/E-TZ/EF) are significantly lower than those for the SF-MLIP without forces (SF-DZ/E), and are of similar accuracy to the SF-MLIP trained directly with DFT/DZ forces (SF-DZ/EF). Although high-level forces would rarely be available without low-level forces, these results provide additional evidence that MFL with forces at one level of theory can achieve high accuracy for forces at a different level of theory that are not included in training. For completeness, Supplementary Material Table~\ref{tbl:SI-DFT/DZ and DFT/TZ} reports errors for all SF and MF MLIPs trained on DFT/DZ and/or DFT/TZ data when tested on both the DFT/DZ and DFT/TZ datasets. 

\begin{table}[htb!]
  \caption{Reverse test. Comparison of energy and force root-mean-square errors (RMSEs) for single-fidelity (SF) and multi-fidelity (MF) MLIPs trained with DFT energies and with/without DFT forces for DZ and TZ basis sets. Errors are computed based on the DFT/DZ dataset. Error bars represent 95\% confidence interval from an ensemble of eight models.}
  \label{tbl:DFT/DZ and DFT/TZ MF-DZ}
  \begin{tabular}{c c c}
  \hline
     & Energy-RMSE (kcal/mol) & Force-RMSE (kcal/mol/\AA) \\
  \hline
SF-DZ/E& 2.91 $\pm$ 0.04 & 13.81 $\pm$ 0.51 \\ 
SF-DZ/EF& 1.40 $\pm$ 0.03 & 3.17 $\pm$ 0.07 \\
MF-DZ/E-TZ/E& 2.84 $\pm$ 0.04& 12.20 $\pm$ 0.09\\
MF-DZ/E-TZ/EF& 1.54 $\pm$ 0.03 & 4.29 $\pm$ 0.13 \\
MF-DZ/EF-TZ/EF& 1.38 $\pm$ 0.02 & 3.45 $\pm$ 0.13 \\
\hline
  \end{tabular}
\end{table}

\begin{table}[htb!]
  \caption{All errors for test case one with tensor order two. Comparison of energy and force errors for single-fidelity (SF) and multi-fidelity (MF) MLIPs trained with DFT energies and with/without DFT forces for DZ and TZ basis sets. Multi-fidelity MLIPs trained with DFT forces from a different basis set achieves nearly the same level of accuracy as single-fidelity MLIPs trained directly to forces with the same basis set. Error bars represent 95\% confidence interval from an ensemble of eight models.}
  \label{tbl:SI-DFT/DZ and DFT/TZ}
  \begin{tabular}{c c c c c}
  \hline
     & \multicolumn{2}{c}{Energy-RMSE (kcal/mol)} & \multicolumn{2}{c}{Force-RMSE (kcal/mol/\AA)} \\

     & DFT/DZ & DFT/TZ & DFT/DZ & DFT/TZ \\
\hline
SF-DZ/E& 2.91 $\pm$ 0.04 & -- & 13.81 $\pm$ 0.51 & -- \\ 
SF-TZ/E& -- & 2.75 $\pm$ 0.02 & -- & 13.52 $\pm$ 0.50\\ 
SF-DZ/EF& 1.40 $\pm$ 0.03& -- & 3.17 $\pm$ 0.07& -- \\
SF-TZ/EF& -- & 1.60 $\pm$ 0.02& -- & 3.52 $\pm$ 0.09\\
MF-DZ/E-TZ/E& 2.84 $\pm$ 0.04& 2.78 $\pm$ 0.04& 12.20 $\pm$ 0.09& 11.92 $\pm$ 0.15\\
MF-DZ/EF-TZ/E& 1.49 $\pm$ 0.04& 1.49 $\pm$ 0.04& 3.95 $\pm$ 0.09& 4.16 $\pm$ 0.15\\
MF-DZ/E-TZ/EF& 1.54 $\pm$ 0.03& 1.47 $\pm$ 0.03& 4.29 $\pm$ 0.13& 3.97 $\pm$ 0.14\\
MF-DZ/EF-TZ/EF& 1.38 $\pm$ 0.02& 1.34 $\pm$ 0.02& 3.45 $\pm$ 0.13& 3.42 $\pm$ 0.11\\
\hline
  \end{tabular}
\end{table}

To provide additional evidence for the performance of MFL, we repeat all of the aforementioned trainings of SF-MLIPs and MF-MLIPs to DFT/DZ and/or DFT/TZ data but with two different tensor orders of HIP-NN. While the results presented in the main text use tensor order of two for HIP-NN, Supplementary Material Table~\ref{tbl:tensor_order_zero} and Supplementary Material Table~\ref{tbl:tensor_order_one} provide results for HIP-NN with tensor orders of zero and one, respectively. Although lower tensor orders naturally have decreased accuracy (i.e., higher energy and force RMSEs), the level of improvement with MFL for each tensor order is quite similar. 

\begin{table}[htb!]
  \caption{All errors for test case one with tensor order zero. Comparison of energy and force errors for HIP-NN tensor order zero for single-fidelity (SF) and multi-fidelity (MF) MLIPs trained with DFT energies and with/without DFT forces for DZ and TZ basis sets. Multi-fidelity MLIPs trained with DFT forces from a different basis set achieves nearly the same level of accuracy as single-fidelity MLIPs trained directly to forces with the same basis set. Errors are reported for a single model.}
  \label{tbl:tensor_order_zero}
  \begin{tabular}{c c c c c}
  \hline
     & \multicolumn{2}{c}{Energy-RMSE (kcal/mol)} & \multicolumn{2}{c}{Force-RMSE (kcal/mol/\AA)} \\

     & DFT/DZ & DFT/TZ & DFT/DZ & DFT/TZ \\
\hline
SF-DZ/E& 5.14 & -- & 25.91 & -- \\ 
SF-TZ/E& -- & 4.85 & -- & 21.35 \\ 
SF-DZ/EF& 2.80 & -- & 7.23 & -- \\
SF-TZ/EF& -- & 2.52 & -- & 6.14 \\
MF-DZ/E-TZ/E& 4.92 & 4.80 & 24.26 & 23.77\\
MF-DZ/EF-TZ/E& 2.80 & 2.76 & 6.97 & 7.20 \\
MF-DZ/E-TZ/EF& 2.88 & 2.78 & 7.28 & 6.78\\
MF-DZ/EF-TZ/EF& 2.62 & 2.56 & 6.30 & 6.17 \\
\hline
  \end{tabular}
\end{table}

\begin{table}[htb!]
  \caption{All errors for test case one with tensor order one. Comparison of energy and force errors for HIP-NN tensor order zero for single-fidelity (SF) and multi-fidelity (MF) MLIPs trained with DFT energies and with/without DFT forces for DZ and TZ basis sets. Multi-fidelity MLIPs trained with DFT forces from a different basis set achieves nearly the same level of accuracy as single-fidelity MLIPs trained directly to forces with the same basis set. Errors are reported for a single model.}
  \label{tbl:tensor_order_one}
  \begin{tabular}{c c c c c}
  \hline
     & \multicolumn{2}{c}{Energy-RMSE (kcal/mol)} & \multicolumn{2}{c}{Force-RMSE (kcal/mol/\AA)} \\

     & DFT/DZ & DFT/TZ & DFT/DZ & DFT/TZ \\
\hline
SF-DZ/E& 4.01 & -- & 19.70 & -- \\ 
SF-TZ/E& -- & 3.82 & -- & 18.26 \\ 
SF-DZ/EF& 1.70 & -- & 4.65 & -- \\
SF-TZ/EF& -- & 1.71 & -- & 4.49 \\
MF-DZ/E-TZ/E& 3.69 & 3.61 & 16.09 & 15.64\\
MF-DZ/EF-TZ/E& 1.95 & 1.97 & 5.21 & 5.60 \\
MF-DZ/E-TZ/EF& 1.90 & 1.82 & 5.06 & 4.66\\
MF-DZ/EF-TZ/EF& 1.69 & 1.65 & 4.15 & 4.08 \\
\hline
  \end{tabular}
\end{table}

\clearpage
\newpage

\section{Further validation of test case two: DFT/DZ and MP2/TZ}

\begin{table}[htb!]
  \caption{All errors for test case two. Comparison of energy and force root-mean-square errors (RMSEs) for single-fidelity (SF) and multi-fidelity (MF) MLIPs trained with either DFT/DZ or MP2/TZ energies with and without DFT/DZ forces. Training to lower level DFT/DZ forces improved the prediction on higher level MP2/TZ energies and forces. No MP2/TZ forces were used during training of SF or MF MLIPs. MP2/TZ force RMSEs are reported for the GDB10to13 testset. Error bars represent 95\% confidence interval from an ensemble of eight models.}
  \label{tbl:SI-DFT/DZ and MP2/TZ}
  \begin{tabular}{c c c c c}
  \hline
     & \multicolumn{2}{c}{Energy-RMSE (kcal/mol)} & \multicolumn{2}{c}{Force-RMSE (kcal/mol/\AA)} \\

     & DFT/TZ & MP2/TZ & DFT/TZ & MP2/TZ (GDB10to13) \\
\hline
SF-DZ/E & 2.91 $\pm$ 0.04 & -- & 13.81 $\pm$ 0.51 & 10.70 $\pm$ 0.27 \\
SF-DZ/EF & 1.40 $\pm$ 0.03 & -- & 3.17 $\pm$ 0.07 & 6.68 $\pm$ 0.03 \\
SF-MP2/E & -- & 2.95 $\pm$ 0.02 & -- & 8.59 $\pm$ 0.42 \\ 
MF-DZ/EF-MP2/E & 1.53 $\pm$ 0.03 & 1.72 $\pm$ 0.02 & 3.91 $\pm$ 0.18 & 3.42 $\pm$ 0.09 \\
\hline
  \end{tabular}
\end{table}

We provide further evidence that MFL is not simply learning the low-level forces. Table~\ref{tbl:SI-GDB10to13 MP2/TZ_testcase2} reports the RMSE for several models relative to the GDB10to13 MP2/TZ forces. Although SF-DZ/EF was trained with the same force data as MF-DZ/EF-MP2/E, the RMSE of SF-DZ/EF on GDB10to13 MP2/TZ forces is nearly twice as high (6.68 kcal/mol/\AA~compared to 3.42 kcal/mol/\AA). Similarly, although the ANI-1x MLIP was trained to DFT/DZ energies (without forces) for the entire ANI-1x dataset consisting of $\approx$4.5M molecular configurations, the RMSE for the ANI-1x MLIP on GDB10to13 MP2/TZ forces is more than a factor of two higher than the MF-DZ/EF-MP2/E force RMSE (9.12 kcal/mol/\AA~compared to 3.42 kcal/mol/\AA). The conclusions from these comparisons are unanimous, MFL is outperforming the accuracy that could be attributed solely to predicting the low-level forces on the high-level output node.


\begin{table}[htb!]
  \caption{Comparison of force MAEs and RMSEs for several MLIPs tested against the MP2/TZ forces in the GDB10to13 dataset. ANI-1x values were computed in this work based on errors for each individual ensemble member. Error bars represent 95\% confidence interval from an ensemble of eight models.}
  \label{tbl:SI-GDB10to13 MP2/TZ_testcase2}
  \begin{tabular}{c c c}
  \hline
  & \multicolumn{2}{c}{MP2/TZ (GDB10to13)}  \\

     & Force-MAE (kcal/mol/A) & Force-RMSE (kcal/mol/A) \\
\hline
SF-DZ/E & 6.90 $\pm$ 0.16 & 10.70 $\pm$ 0.27 \\
SF-MP2/E & 5.55 $\pm$ 0.21 & 8.59 $\pm$ 0.42 \\  
SF-DZ/EF & 4.15 $\pm$ 0.02 & 6.68 $\pm$ 0.03 \\
MF-DZ/EF-MP2/E & 2.23 $\pm$ 0.05 & 3.42 $\pm$ 0.09 \\
ANI-1x & 5.84 $\pm$ 0.09 & 9.12 $\pm$ 0.11 \\
\hline
  \end{tabular}
\end{table}

\clearpage
\newpage

\section{Further validation of test case three: DFT/DZ and CCSD(T)*/CBS}

Supplementary Material Tables~\ref{tbl:SI-DFT/DZ and CC}-\ref{tbl:SI-GDB10to13 MP2/TZ_testcase3} provide further evidence that MFL with low-level forces significantly outperforms SF with only high-level energies. By contrast, MFL without forces (only energies for both levels of theory) provides marginal improvement, consistent with the results of Jacobson et al. and the first test case. Specifically, the energy, force, $\Delta E_{\rm conf}$, and $\Delta E_{\rm tors}$-RMSEs for MF-DZ/E-CC/E are statistically indistinguishable from the respective RMSEs for SF-CC/E, whereas the RMSEs for MF-DZ/EF-CC/E are nearly a factor of two lower. 

\begin{table}[htb!]
      \caption{All errors for test case three. Comparison of energy and force root-mean-square errors (RMSEs) for single-fidelity (SF) and multi-fidelity (MF) MLIPs trained with either DFT/DZ or CCSD(T)*/CBS energies with and without DFT/DZ forces. Training to lower level DFT/DZ forces improved the prediction on higher level CCSD(T)*/CBS energies and MP2/TZ forces. No CCSD(T)*/CBS forces were used during training of SF or MF MLIPs. CCSD(T)*/CBS energy RMSEs are reported for the reduced ANI-1x dataset. MP2/TZ force RMSEs are reported for the GDB10to13 testset. Error bars represent 95\% confidence interval from an ensemble of eight models.}
      \label{tbl:SI-DFT/DZ and CC}
      \begin{tabular}{c c c c c}
      \hline
         & \multicolumn{2}{c}{Energy-RMSE (kcal/mol)} & \multicolumn{2}{c}{Force-RMSE (kcal/mol/A)} \\

         & DFT/DZ & CCSD(T)*/CBS & DFT/DZ & MP2/TZ (GDB10to13)\\
    \hline
    SF-DZ/E & 2.91 $\pm$ 0.04 & -- & 13.81 $\pm$ 0.51 & 10.70 $\pm$ 0.27 \\ 
    SF-CC/E & -- & 2.78 $\pm$ 0.05 & -- & 8.85 $\pm$ 0.14 \\ 
    SF-DZ/EF & 1.40 $\pm$ 0.03 & -- & 3.17 $\pm$ 0.07 & 6.68 $\pm$ 0.03 \\
    MF-DZ/E-CC/E & 2.90 $\pm$ 0.03 & 2.76 $\pm$ 0.03 & 13.58 $\pm$ 0.51 & 8.76 $\pm$ 0.24 \\
    MF-DZ/EF-CC/E & 1.54 $\pm$ 0.03 & 1.52 $\pm$ 0.03 & 3.76 $\pm$ 0.11 & 4.86 $\pm$ 0.06 \\
    \hline
      \end{tabular}
\end{table}

Although it is not possible to compare absolute energies for MLIPs trained to different QM levels of theory, due to different energy reference points, a comparison of energy differences $(\Delta E)$ is more meaningful. For this reason, Supplementary Material Table \ref{tbl:SI-DeltaE CC} compares the CC-level $\Delta E_{\rm conf}$-RMSEs and $\Delta E_{\rm conf}$-MAEs for HIP-NN and ANI MLIPs trained to DFT/DZ, DFT/TZ, and/or CCSD(T)*/CBS data. 


\begin{table}[htb!]
      \caption{Comparison of conformer energy MAEs and RMSEs for several MLIPs tested against the CCSD(T)*/CBS $\Delta E_{\rm conf}$ values in the GDB10to13 dataset. ANI values were computed in this work based on errors for each individual ensemble member. Error bars represent 95\% confidence interval from an ensemble of eight models.}
      \label{tbl:SI-DeltaE CC}
      \begin{tabular}{c c c}
      \hline 
          Model & $\Delta E_{\rm conf}$-MAE (kcal/mol) & $\Delta E_{\rm conf}$-RMSE (kcal/mol) \\
    \hline
    SF-DZ/E & 2.41 $\pm$ 0.04 & 3.37 $\pm$ 0.05 \\
    SF-DZ/EF & 1.72 $\pm$ 0.02 & 2.44 $\pm$ 0.03 \\
    SF-CC/E & 1.90 $\pm$ 0.03 & 2.65 $\pm$ 0.04 \\ 
    MF-DZ/E-CC/E & 1.87 $\pm$ 0.02 & 2.60 $\pm$ 0.03 \\
    MF-DZ/EF-CC/E & 1.13 $\pm$ 0.02 & 1.60 $\pm$ 0.03 \\
    MF-TZ/E-CC/E & 1.93 $\pm$ 0.03 & 2.69 $\pm$ 0.05 \\
    MF-TZ/EF-CC/E & 1.14 $\pm$ 0.01 & 1.61 $\pm$ 0.01 \\
    MF-DZ/EF-TZ/EF-CC/E & 1.38 $\pm$ 0.02 & 1.96 $\pm$ 0.03 \\
    ANI-1x & 2.32 $\pm$ 0.03  & 3.26 $\pm$ 0.05 \\
    ANI-1ccx & 1.88 $\pm$ 0.02 & 2.62 $\pm$ 0.03 \\
    ANI-1ccx-R & 2.39 $\pm$ 0.02 & 3.31 $\pm$ 0.04 \\
    \hline
      \end{tabular}
\end{table}

Supplementary Material Table \ref{tbl:SI-DeltaE tors} also compares the CCSD(T)*/CBS-level $\Delta E_{\rm tors}$-RMSEs and $\Delta E_{\rm tors}$-MAEs for HIP-NN and ANI MLIPs trained to the CCSD(T)*/CBS data. All MLIPs were evaluated using the same geometries provided by Sellers et al., i.e., without performing a relaxed torsion scan.


\begin{table}[htb!]
      \caption{Comparison of torsion scan energy MAEs and RMSEs for several MLIPs tested against the CCSD(T)*/CBS $\Delta E_{\rm tors}$ values in the torsion scan dataset. ANI values were computed in this work based on errors for each individual ensemble member. Error bars represent 95\% confidence interval from an ensemble of eight models.}
      \label{tbl:SI-DeltaE tors}
      \begin{tabular}{c c c}
      \hline 
          Model & $\Delta E_{\rm tors}$-MAE (kcal/mol) & $\Delta E_{\rm tors}$-RMSE (kcal/mol) \\
    \hline
    SF-CC/E & 0.51 $\pm$ 0.03 & 0.71 $\pm$ 0.04 \\ 
    MF-DZ/E-CC/E & 0.48 $\pm$ 0.03 & 0.67 $\pm$ 0.04 \\
    MF-DZ/EF-CC/E & 0.28 $\pm$ 0.01 & 0.40 $\pm$ 0.02 \\
    ANI-1ccx & 0.54 $\pm$ 0.05 & 0.77 $\pm$ 0.07 \\
    ANI-1ccx-R & 0.66 $\pm$ 0.05 & 0.93 $\pm$ 0.07 \\ 
    \hline
      \end{tabular}
\end{table}

As neither the ANI-1ccx or GDB10to13 datasets contain CC-level forces, we compare our HIP-NN models and ANI models with CC-level forces from the Allen et al. dataset~\cite{allen_reactive_2025} and MP2/TZ forces from GDB10to13. 

The complete Allen et al. dataset consists of approximately 1000 configurations for reactants, products, and transition states. However, because the ANI-1ccx dataset does not contain chemical reactions, we filter out all transition states. Furthermore, because many of the reactants and products are not well-represented by the ANI-1ccx dataset, we also filter out all structures where the normalized energy standard deviation for the ANI-1ccx ensemble is greater than the active learning threshold (0.23 kcal/mol/$\sqrt N$) used by Smith et al. We test each MLIP only on this reduced dataset consisting of approximately 100 reactants and products. 

The CC-level forces in Allen et al. are computed with a different protocol than the CC-level energies in the ANI-1ccx dataset. Specifically, Allen et al. employed a basis set correction methodology to approximate quadruple zeta (referred to as QZ*) that is different from the complete basis set extrapolation approach of Smith et al. Allen et al. report an RMSE between forces computed with QZ* and true QZ of 1.7 kcal/mol/\AA. The difference between QZ* and CBS should be slightly larger than this value, since even true QZ has not fully converged to the CBS limit. Another difference is that Allen et al. used unrestricted calculations, whereas Smith et al. used restricted calculations. Allen et al. reports a force RMSE between UCCSD(T)/DZ* and RCCSD(T)/DZ of 2.0 kcal/mol/\AA. Considering the accumulation of errors caused by the two differences mentioned here, some caution should be exercised when comparing the UCCSD(T)/QZ* forces of Allen et al. with MLIPs trained to the RCCSD(T)*/CBS energies of ANI-1ccx.

Supplementary Material Table~\ref{tbl:SI-Allen_forces} reports the MAEs and RMSEs between the CC-level forces for our reactants/products subset of the Allen et al. dataset compared with several MLIPs trained with the CC-level energies of ANI-1ccx. The MF-DZ/EF-CC/E model again achieves RMSEs (2.69 kcal/mol/\AA) significantly lower than SF-CC/E (4.84 kcal/mol/\AA), MF-DZ/E-CC/E (5.10~kcal/mol/\AA) and ANI-1ccx (4.60 kcal/mol/\AA), providing definitive evidence that MFL with low-level forces improves the accuracy on high-level forces.

\begin{table}[htb!]
  \caption{Comparison of force MAEs and RMSEs for several MLIPs tested against the UCCSD(T)/QZ* forces for a reduced reactants/products subset of the Allen et al. dataset. ANI values were computed in this work based on errors for each individual ensemble member. Error bars represent 95\% confidence interval from an ensemble of eight models.}
  \label{tbl:SI-Allen_forces}
  \begin{tabular}{c c c}
  \hline
  & \multicolumn{2}{c}{UCCSD(T)/QZ* (Allen et al.)}  \\

  & Force-MAE (kcal/mol/\AA) & Force-RMSE (kcal/mol/\AA) \\
\hline
SF-CC/E & 3.15 $\pm$ 0.08 & 4.84 $\pm$ 0.13\\ 
MF-DZ/E-CC/E & 3.27 $\pm$ 0.17 & 5.10 $\pm$ 0.30 \\
MF-DZ/EF-CC/E & 1.71 $\pm$ 0.02 & 2.69 $\pm$ 0.04\\
ANI-1ccx & 2.98 $\pm$ 0.09 & 4.60 $\pm$ 0.18\\
ANI-1ccx-R & 3.30 $\pm$ 0.10 & 5.10 $\pm$ 0.16 \\
\hline
  \end{tabular}
\end{table}

Supplementary Material Table~\ref{tbl:SI-GDB10to13 MP2/TZ_testcase3} provides further evidence that MFL improves the high-level forces by comparing with the GDB10to13 MP2/TZ forces. Specifically, we compare the RMSE and MAE for the single-fidelity MLIP trained just to CCSD(T)*/CBS energies (SF-CC/E) and the multi-fidelity MLIP trained to DFT/DZ energies and forces and CCSD(T)*/CBS energies (MF-DZ/EF-CC/E). The MF-DZ/EF-CC/E errors are considerably lower than the SF-CC/E errors on the MP2/TZ forces (4.86 kcal/mol/\AA~vs 8.85 kcal/mol/\AA). Although still impressive, the improvement in the force RMSE for MF-DZ/EF-CC/E compared to SF-CC/E is not as pronounced as the first two test cases. We attribute the relatively higher force error to the fact that the RMSEs are computed with respect to MP2/TZ forces while the MLIP was trained to CCSD(T)*/CBS energies. Nonetheless, the MF-DZ/EF-CC/E RMSE (4.86 kcal/mol/\AA) is still considerably lower than the RMSE for the ANI-1ccx MLIP (7.27 kcal/mol/\AA) tested on the GDB10to13 MP2/TZ forces.



\begin{table}[htb!]
  \caption{Comparison of force MAEs and RMSEs for several MLIPs tested against the MP2/TZ forces in the GDB10to13 dataset. ANI values were computed in this work based on errors for each individual ensemble member. Error bars represent 95\% confidence interval from an ensemble of eight models.}
  \label{tbl:SI-GDB10to13 MP2/TZ_testcase3}
  \begin{tabular}{c c c}
  \hline
  & \multicolumn{2}{c}{MP2/TZ (GDB10to13)}  \\

     & Force-MAE (kcal/mol/\AA) & Force-RMSE (kcal/mol/\AA) \\
\hline
SF-DZ/E & 6.90 $\pm$ 0.16 & 10.70 $\pm$ 0.27 \\
SF-CC/E& 5.66 $\pm$ 0.05& 8.85 $\pm$ 0.14\\ 
SF-DZ/EF & 4.15 $\pm$ 0.02 & 6.68 $\pm$ 0.03 \\
MF-DZ/E-CC/E & 5.62 $\pm$ 0.14 & 8.76 $\pm$ 0.24 \\
MF-DZ/EF-CC/E& 2.93 $\pm$ 0.05& 4.86 $\pm$ 0.06\\
ANI-1ccx& 4.68 $\pm$ 0.07 & 7.27 $\pm$ 0.11\\
ANI-1ccx-R & 5.58 $\pm$ 0.05 & 8.47 $\pm$ 0.08\\
\hline
  \end{tabular}
\end{table}

The errors reported in this study for both HIP-NN and ANI models are the average RMSE and average MAE computed from the individual errors for each ensemble member. By contrast, the RMSEs in Smith et al. were computed relative to the ensemble-averaged energies and forces.~\cite{smith2019approaching} For this reason, the RMSEs and MAEs for ANI models reported in Supplementary Material Table~\ref{tbl:SI-DeltaE CC} and Supplementary Material Table~\ref{tbl:SI-GDB10to13 MP2/TZ_testcase3} are higher than those reported by Smith et al. For example, Smith et al. reported RMSEs for ANI-1ccx of 2.07 kcal/mol and 5.3 kcal/mol/\AA~for $\Delta E_{\rm conf}$ and GDB10to13 forces, respectively. For comparison, the MF-DZ/EF-CC/E RMSEs for $\Delta E_{\rm conf}$ and GDB10to13 forces are even lower (1.26 kcal/mol and 4.30 kcal/mol/\AA, respectively) when the errors are computed relative to ensemble-averaged energies and forces. Furthermore, when computing $\Delta E_{\rm tors}$ with ensemble-averaged energies and using the Sellers-provided geometries, the RMSE for MF-DZ/EF-CC/E of 0.33 kcal/mol is also significantly lower than the RMSE of 0.52 kcal/mol for ANI-1ccx (calculated in this study).

\begin{figure}[htb!]
    \includegraphics[width=3in]{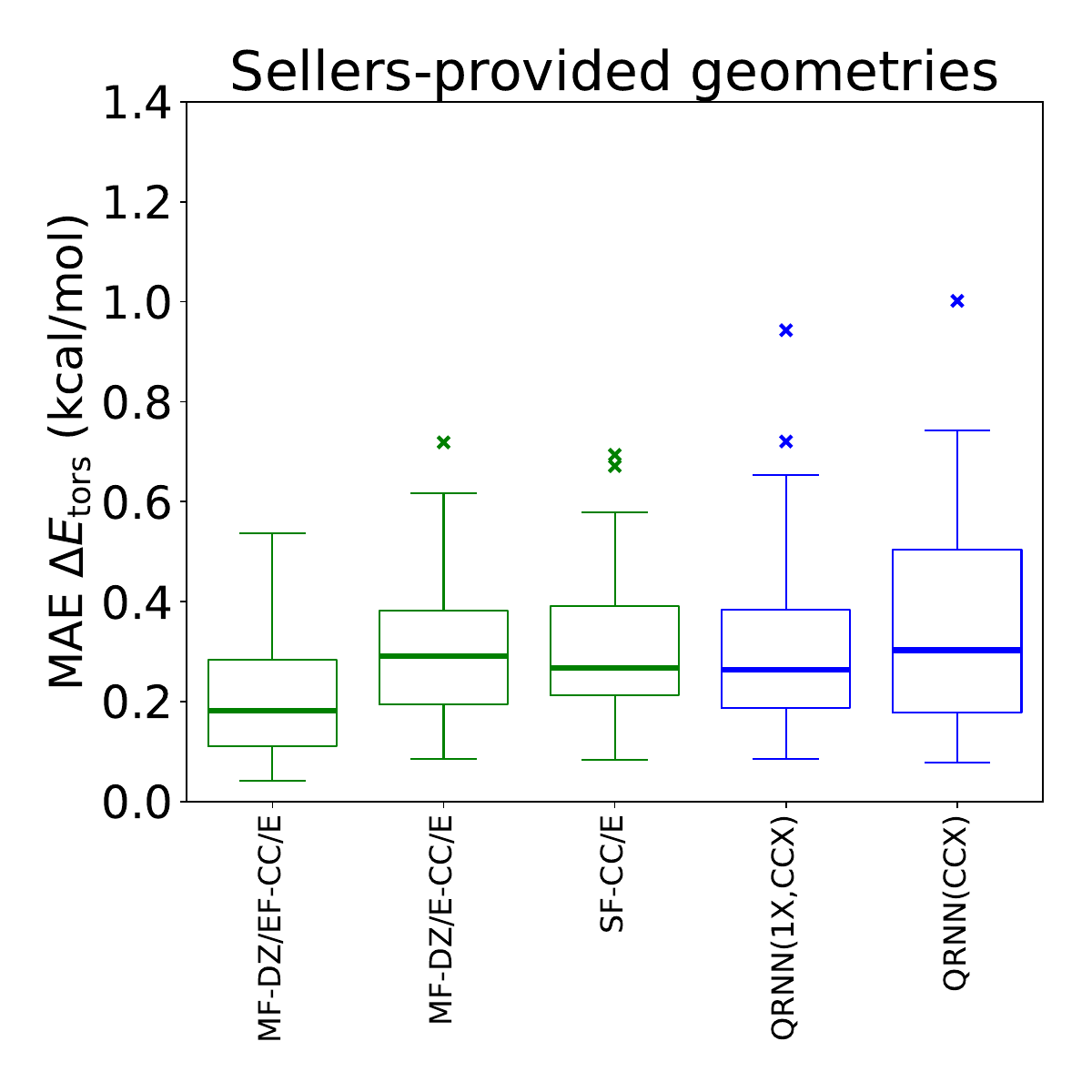}
    \includegraphics[width=3in]{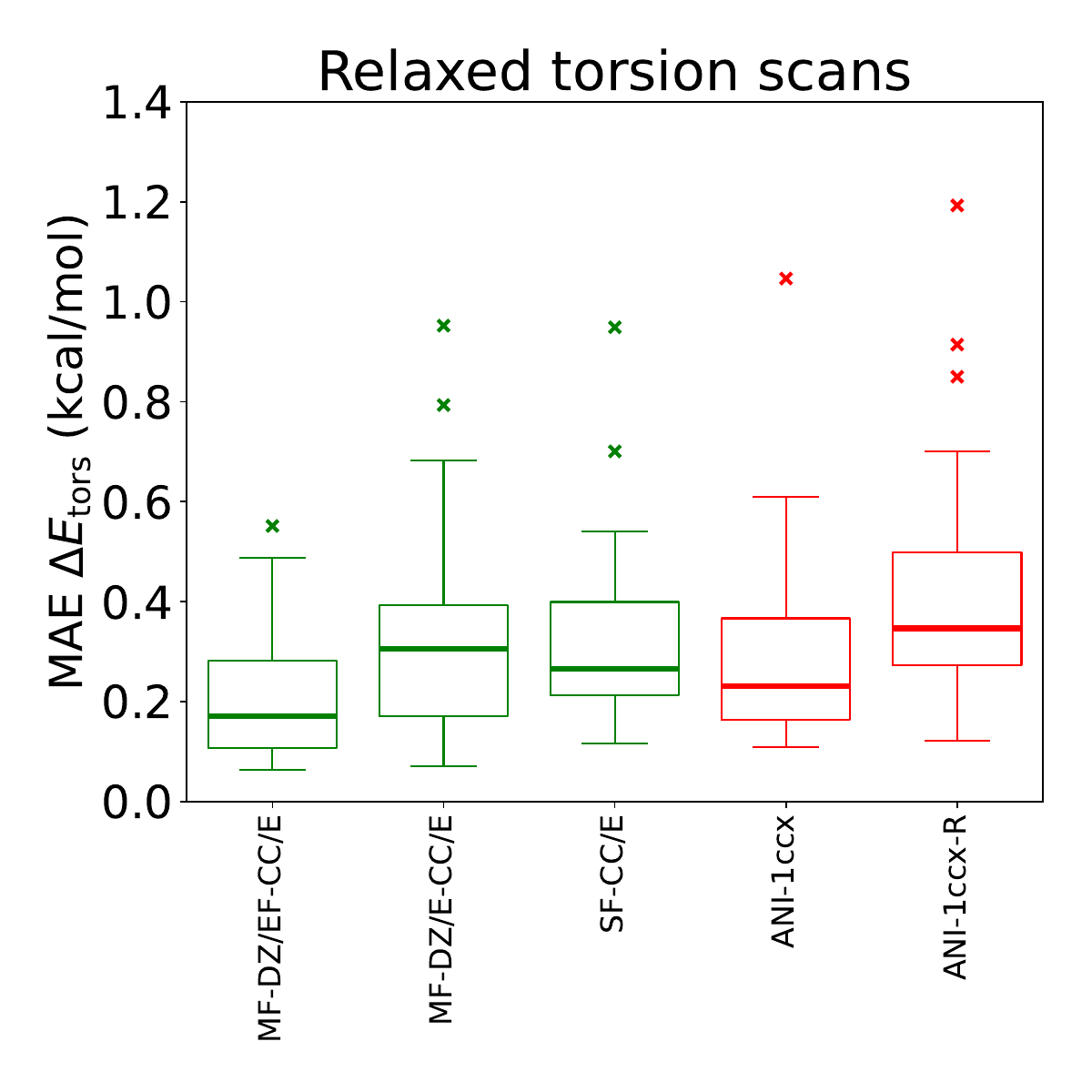}
    \caption{Comparison of MAEs for SF and MF HIP-NN models in this work and ANI models from the literature. QRNN(CCX) and QRNN(1X,CCX) are the ANI models of Jacobson et al. with SF and MF training, respectively. ANI-1ccx and ANI-1ccx-R are the ANI models of Smith et al. with and without transfer learning, respectively. The results in the left panel were obtained using the provided geometries of Sellers et al. The results in the right panel were obtained following a relaxed torsion scan, wherein the positions of all atoms are optimized with the corresponding MLIP subject to the constraint of a fixed dihedral angle. The box extends from the upper to lower quartile. The horizontal line in the box is the median. The upper ``whisker'' extends to the last data point less than the third quartile plus 1.5 times the interquartile range while the lower ``whisker'' extends to the first data point greater than the first quartile minus 1.5 times the interquartile range. For clarity, only points beyond the ``whiskers'' are represented with symbols.}
    \label{fig:SI-torsion_scan}
\end{figure}

Figure~\ref{fig:SI-torsion_scan} provides a comparison of several models on the torsion scan test dataset. To compare with previous studies, we utilize the CCSD(T)/CBS energies reported by Sellers et al. Consistent with Jacobson et al. and Smith et al., we also utilize the ensemble-averaged energies to compute the MAEs for the SF and MF HIP-NN models. Consistent with these studies, a separate MAE is computed for each individual molecule and then the median and quartiles are computed based on the 45 molecule-specific MAEs. As Smith et al. performed a relaxed torsion scan while Jacobson et al. used the geometries provided in the dataset, we report our results using both approaches. The small difference in these techniques demonstrates that the relaxed geometries are quite similar to the original geometries of Sellers et al., confirming that our HIP-NN models can perform torsion scans without requiring pre-relaxed geometries. Notably, MF-DZ/EF-CC/E achieves the lowest median MAE of 0.17 kcal/mol and 0.18 kcal/mol for the relaxed torsion scan and Sellers-provided geometries, respectively.

\clearpage
\newpage

Supplementary Material Table~\ref{tbl:SI-DeltaE CC} and Supplementary Material Table~\ref{tbl:SI-Triple multi-fidelity} demonstrate that using slightly more accurate low-level (DFT/TZ) forces does not significantly improve the performance of MFL. Specifically, the energy and $\Delta E_{\rm conf}$-RMSE for MF-TZ/EF-CC/E are statistically indistinguishable from the RMSEs for MF-DZ/EF-CC/E. 

We also test MFL with energies and forces for both DFT/DZ and DFT/TZ along with the CCSD(T)*/CBS energies. Notably, the CC-level errors are actually considerably worse for MFL with three datasets than MFL with just two datasets (see Supplementary Material Table~\ref{tbl:SI-Triple multi-fidelity}). Specifically, the energy, $\Delta E_{\rm conf}$, and force RMSEs for MF-DZ/EF-TZ/EF-CC/E are all markedly higher (1.61 kcal/mol, 1.96 kcal/mol, and 5.27 kcal/mol/\AA) than those for either MF-DZ/EF-CC/E (1.52 kcal/mol, 1.60 kcal/mol, and 4.86 kcal/mol/\AA) or MF-TZ/EF-CC/E (1.49 kcal/mol, 1.61 kcal/mol, and 4.88 kcal/mol/\AA), respectively.

\begin{sidewaystable}[htb!]
  \caption{Multi-fidelity learning to DFT/DZ, DFT/TZ, and CCSD(T)*/CBS does not improve compared to multi-fidelity with just DFT and CC. Error bars represent 95\% confidence interval from an ensemble of eight models.}
  \label{tbl:SI-Triple multi-fidelity}
  \begin{tabular}{c c c c c c c}
  \hline
     & \multicolumn{3}{c}{Energy-RMSE (kcal/mol)} & \multicolumn{3}{c}{Force-RMSE (kcal/mol/A)} \\

     & DFT/DZ & DFT/TZ & CC & DFT/DZ & DFT/TZ & MP2/TZ (GDB10to13) \\
\hline
SF-DZ/E & 2.91 $\pm$ 0.04 & -- & -- & 13.81 $\pm$ 0.51 & -- & 12.53 $\pm$ 0.40 \\ 
SF-TZ/E & -- & 2.75 $\pm$ 0.02 & -- & -- & 13.52 $\pm$ 0.50 & 11.83 $\pm$ 0.28 \\ 
MF-DZ/EF-CC/E & 1.54 $\pm$ 0.03 & -- & 1.52 $\pm$ 0.03 & 3.76 $\pm$ 0.11 & -- & 4.86 $\pm$ 0.06 \\
MF-TZ/EF-CC/E & -- & 1.47 $\pm$ 0.02 & 1.49 $\pm$ 0.03 & -- & 3.90 $\pm$ 0.10 & 4.88 $\pm$ 0.03 \\
MF-DZ/EF-TZ/EF-CC/E & 1.64 $\pm$ 0.04 & 1.60 $\pm$ 0.04 & 1.61 $\pm$ 0.03 & 3.96 $\pm$ 0.19 & 3.92 $\pm$ 0.16 & 5.27 $\pm$ 0.03 \\
\hline
  \end{tabular}
\end{sidewaystable}

\clearpage
\newpage

\section{Further percent improvements}

\begin{figure}[htb!]
    \includegraphics[width=3in]{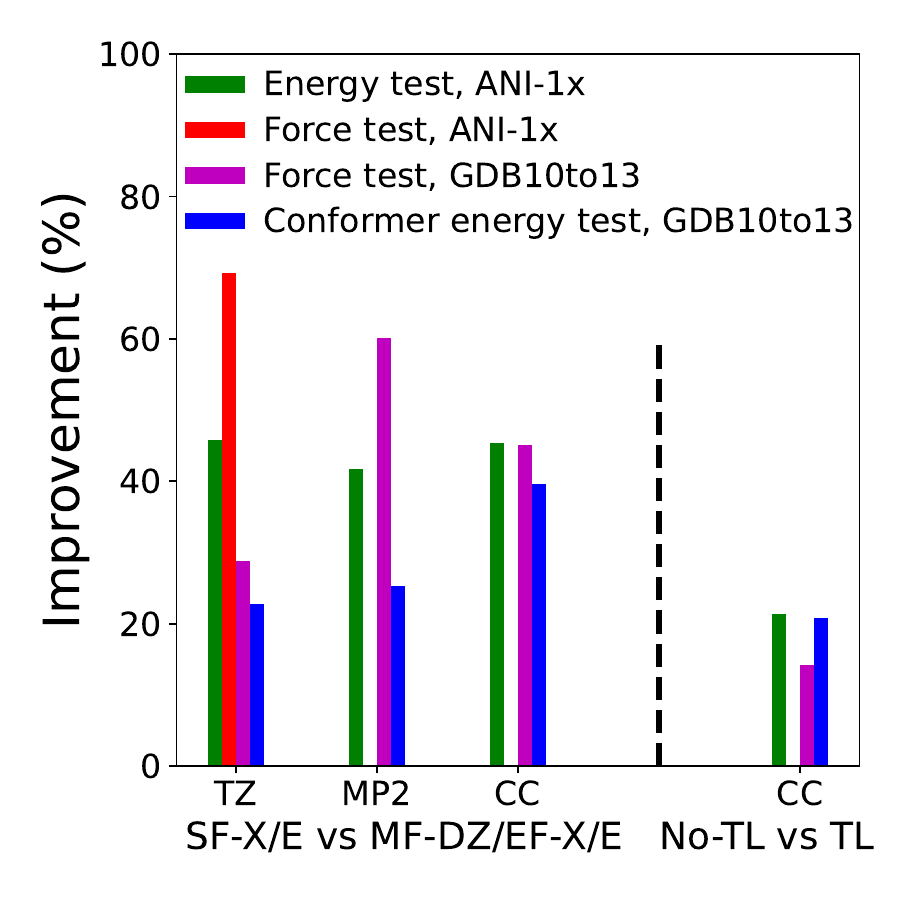}
    \caption{Additional comparisons of percent improvement for multi-fidelity learning (MFL) with three test cases and for transfer learning (TL) with ANI-1ccx. Percent improvements are computed for the energy, forces, and conformer energies $(\Delta E_{\rm conf})$. Force improvement is computed for both the in-sample test dataset (held-out data from ANI-1x) and for the out-of-sample GDB10to13 test dataset. `X' represents any of the different fidelities, namely, DFT/TZ, MP2/TZ, and CCSD(T)*/CBS. GDB10to13 comparisons are computed with the MP2/TZ forces and the CCSD(T)*/CBS conformer energy differences $(\Delta E_{\rm conf})$. Improvement percentage for MFL is calculated by $($RMSE$_{\rm SF}-$RMSE$_{\rm MF})/$RMSE$_{\rm SF} \times 100\%$ while improvement percentage for TL is calculated by $($RMSE$_{\rm no TL}-$RMSE$_{\rm TL})/$RMSE$_{\rm no TL} \times 100\%$, where ``no TL'' corresponds to the ANI-1ccx-R MLIP.}
    \label{fig:SI-Improvement factor rough chart}
\end{figure}

Figure~\ref{fig:SI-Improvement factor rough chart} provides additional percent improvements. Specifically, the percent improvements for the MF-MLIPs trained to TZ-level and CC-level data are also tested on GDB10to13 MP2/TZ forces and CCSD(T)*/CBS $\Delta E_{\rm conf}$s. The improvement on the GDB10to13 MP2/TZ forces spans a range of 30\% for MF-DZ/EF-TZ/E, 45\% for MF-DZ/EF-CC/E, and 60\% for MF-DZ/EF-MP2/E. The two lowest percent improvements (MF-DZ/EF-TZ/E and MF-DZ/EF-CC/E) were for cases where the MLIP was not actually trained with energies at the same level of theory as the GDB10to13 forces (MP2/TZ). Thus, when the MLIP predicts high-level forces that exactly correspond to the same level of theory as the test forces, the percent improvement in forces is between 60\%-70\%. Similarly, although the improvement in conformer energies ($\Delta E_{\rm conf}$) ranges from 20\%-40\% for the three MFL test cases, the highest improvement occurs for MF-DZ/EF-CC/E, which is also when the high-level training energies correspond to the same level of theory (CCSD(T)*/CBS) as the GDB10to13 conformer energies. Table~\ref{tbl:All comparison} provides the RMSEs used to compute all percent improvements.

\begin{sidewaystable}[htb!]
  \caption{Comparison of root-mean-square errors (RMSEs) between single-fidelity and multi-fidelity models for all three test cases. Error bars represent 95\% confidence interval from an ensemble of eight models.}
  \label{tbl:All comparison}
  \begin{tabular}{c c c c c}
  \hline
     & Energy-RMSE & Force-RMSE (ANI-1x) & Force-RMSE (GDB10to13 MP2/TZ) & $\Delta E_{\rm conf}$-RMSE (GDB10to13 CC) \\
     & (kcal/mol) & (kcal/mol/\AA) & (kcal/mol/\AA) & (kcal/mol) \\
  \hline
SF-TZ/E & 2.75 $\pm$ 0.02 & 13.52 $\pm$ 0.50 & -- & 3.60 $\pm$ 0.02 \\ 
MF-DZ/EF-TZ/E &  1.49 $\pm$ 0.04 & 4.16 $\pm$ 0.15 & -- & 2.78 $\pm$ 0.04 \\
SF-MP2/E & 2.95 $\pm$ 0.02 & -- & 8.59 $\pm$ 0.42 & 2.65 $\pm$ 0.02\\ 
MF-DZ/EF-MP2/E & 1.72 $\pm$ 0.02 & -- & 3.42 $\pm$ 0.09 & 1.98 $\pm$ 0.02 \\
SF-CC/E & 2.78 $\pm$ 0.05 & -- & 8.85 $\pm$ 0.14 & 2.65 $\pm$ 0.04 \\ 
MF-DZ/EF-CC/E & 1.52 $\pm$ 0.03 & -- & 4.86 $\pm$ 0.06 & 1.60 $\pm$ 0.03 \\
\hline
  \end{tabular}
\end{sidewaystable}

\clearpage
\newpage

\section{Further comparison of HIP-NN and ANI}

All errors for ANI models in this work are computed using torchani and the publicly available models for ANI-1x and ANI-1ccx. However, because the ANI-1ccx-R model is not publicly available, we retrained a model in torchani implementing the same hyperparameters and learning protocol without using transfer learning. The ANI errors reported herein are different than those reported by Smith et al. because our errors are not computed relative to the ensemble-averaged energies and forces. To confirm that our implementation and analysis are consistent with Smith et al., we verify that, when computed relative to ensemble-averaged energies and forces, we achieve identical errors to those reported in Smith et al. for ANI-1x and ANI-1ccx (and similar errors for ANI-1ccx-R). 

Although MF-DZ/EF-CC/E is unequivocally more accurate than the ANI-1ccx MLIP, it is important to exercise some caution when comparing the RMSEs for these two models. Not only is ANI-1ccx based on a different MLIP architecture, but ANI-1ccx was pre-trained with a slightly different training dataset. Specifically, the ANI-1ccx MLIP was developed by first training to the DFT/DZ energies (without forces) for all $\approx$4.5M molecular configurations in the ANI-1x dataset, followed by retraining with transfer learning to the CCSD(T)*/CBS energies for the same $\approx$460k molecular configurations in the reduced ANI-1x dataset (a.k.a., the ANI-1ccx dataset). Therefore, it is difficult to elucidate whether the higher accuracy of MF-DZ/EF-CC/E is truly due to MFL outperforming TL or to the difference in MLIP architectures or to the difference in training data. 

Based on previous experience, HIP-NN (with tensor order of two) typically achieves similar or even slightly lower errors than ANI models. Therefore, it is important to consider whether the lower RMSEs for MF-DZ/EF-CC/E relative to ANI-1ccx are attributed to the HIP-NN architecture simply outperforming the ANI architecture. To investigate this possibility, we compare SF-CC/E with the ANI-1ccx-R MLIP, which was trained (without transfer learning) to the exact same dataset of $\approx$460k CCSD(T)*/CBS energies. Notably, the SF-CC/E energy RMSE is substantially lower than that for ANI-1ccx-R (2.78 kcal/mol vs 3.27 kcal/mol), while the SF-CC/E force RMSE (relative to GDB10to13 MP2/TZ forces) is similar to ANI-1ccx-R (8.85 kcal/mol/\AA~vs 8.47 kcal/mol/\AA), and the $\Delta E_{\rm conf}$-RMSE for SF-CC/E is also substantially lower than ANI-1ccx-R (2.65 kcal/mol and 3.31 kcal/mol), respectively. 


To further investigate whether HIP-NN (tensor order of two) simply outperforms ANI, we compare SF-DZ/E with the ANI-1x MLIP. A somewhat surprising result is that the RMSEs for energy (relative to DFT/DZ), force (relative to GDB10to13 MP2/TZ forces), and $\Delta E_{\rm conf}$ (relative to CCSD(T)*/CBS) for SF-DZ/E (2.91 kcal/mol, 10.70 kcal/mol/\AA~and 3.37 kcal/mol) are actually slightly higher than the RMSEs for the ANI-1x MLIP (2.59 kcal/mol, 9.12 kcal/mol/\AA~and 3.26 kcal/mol), respectively (see Supplementary Material Table~\ref{tbl:SI-DFT/DZ and CC}, Supplementary Material Table~\ref{tbl:SI-GDB10to13 MP2/TZ_testcase2} and Supplementary Material Table~\ref{tbl:SI-DeltaE CC}). However, we attribute the poorer performance of SF-DZ/E to the difference in training dataset, not the difference in MLIP architecture. ANI-1x was trained to the DFT/DZ energies (without forces) for all $\approx$4.5M molecular configurations in the ANI-1x dataset, whereas SF-DZ/E was trained to the DFT/DZ energies for only the $\approx$460k molecular configurations in the reduced ANI-1x dataset. Including the DFT/DZ forces for these $\approx$460k configurations gives much lower RMSEs for SF-DZ/EF (1.40 kcal/mol, 6.68 kcal/mol/\AA~and 2.44 kcal/mol). In fact, the energy and force RMSEs for SF-DZ/EF (1.40 kcal/mol and 3.17 kcal/mol/\AA) are similar to the values reported for an alternative ANI MLIP (1.38 kcal/mol and 2.78 kcal/mol/\AA) that was trained with DFT/DZ energy and force data for the entire ANI-1x dataset.~\cite{Smith2020SimpleAE} Overall, these results suggest that the superior performance of MF-DZ/EF-CC/E compared to ANI-1ccx is not primarily attributed to the choice of MLIP architecture, but rather to MFL.

\clearpage
\newpage


\printbibliography